\documentclass[ preprint]{revtex4}
\usepackage{amsmath}
\usepackage{graphicx}

\providecommand{\tabularnewline}{\\}

\usepackage{bm}

\begin{document}

\title{$m=1$ Ideal Internal Kink Modes in a Line-tied Screw Pinch}

\author{Yi-Min Huang,$^{1,2}$ Ellen G. Zweibel,$^{1,2,3}$ and Carl R. Sovinec$^{1,4}$}

\affiliation{$^{1}$Center for Magnetic Self-Organization in Laboratory and Astrophysical
Plasmas, University of Wisconsin, Madison, Wisconsin 53706\\
$^{2}$Department of Physics, University of Wisconsin, Madison, Wisconsin
53706\\
$^{3}$Department of Astronomy, University of Wisconsin, Madison,
Wisconsin 53706\\
$^{4}$Department of Engineering Physics, University of Wisconsin,
Madison, Wisconsin 53706}

\begin{abstract}
It is well known that the radial displacement of the $m=1$ internal
kink mode in a periodic screw pinch has a steep jump at the resonant
surface where $\mathbf{k}\cdot\mathbf{B}=0$.\cite{RosenbluthDR1973}
In a line-tied system, relevant to solar and astrophysical plasmas,
the resonant surface is no longer a valid concept. It is then of interest
to see how line-tying alters the aforementioned result for a periodic
system. If the line-tied kink also produces a steep gradient, corresponding
to a thin current layer, it may lead to strong resistive effects even
with weak dissipation. Numerical solution of the eigenmode equations
shows that the fastest growing kink mode in a line-tied system still
possesses a jump in the radial displacement at the location coincident
with the resonant surface of the fastest growing mode in the periodic
counterpart. However, line-tying thickens the inner layer and slows
down the growth rate. As the system length $L$ approaches infinity,
both the inner layer thickness and the growth rate approach the periodic
values. In the limit of small $\epsilon\sim B_{\phi}/B_{z}$, the
critical length for instability $L_{c}\sim\epsilon^{-3}$. The relative
increase in the inner layer thickness due to line-tying scales as
$\epsilon^{-1}(L_{c}/L)^{2.5}$. 

\end{abstract}
\maketitle

\section{Introduction}

Magnetohydrodynamic (MHD) kink instability in a screw pinch is of
great interest and has been under intensive study for decades. In
a periodic system the internal kink mode is unstable if the safety
factor $q$ (for a cylinder with axial periodicity length $L$, $q\equiv2\pi rB_{z}/LB_{\phi}$)
drops below unity somewhere within the plasma and increases with radius.
Unstable modes have helical symmetry, depending on $z$ and $\phi$
in the combination $kz\pm\phi$, and possess so-called resonant surfaces
$r=r_{s}$, on which $\mathbf{k}\cdot\mathbf{B}=kB_{z}\pm B_{\phi}/r_{s}=0$;
due to the periodicity, the resonant surfaces must also be rational
surfaces (if $k=2n\pi/L$, then on $r=r_{s}$, $q=\pm1/n$ is a rational
number). The driver of the instability lies within $r_{s}$, and the
radial component of the plasma displacement nearly vanishes outside
it. It has been shown that the thin transition layer near $r_{s}$
predicted from linear theory corresponds to an infinite current sheet
in finite amplitude theory, at least within the framework of reduced,
ideal MHD.\cite{RosenbluthDR1973} In a plasma with large but finite
Lundquist number, the steepening of the current layer must trigger
resistive energy release, and it has been suggested that this energy
release corresponds to the sawtooth crash.

The kink instability has also been proposed as a cause of solar flares.
In this scenario, the instability occurs in a force free coronal magnetic
loop which emerges from the photosphere and is twisted by photospheric
motions. The possibility of forming a thin current sheet via the kink
mode is particularly interesting, because the Lundquist number $S$
in coronal loops is so high ($\sim10^{10-12}$) as to otherwise effectively
preclude fast resistive MHD processes. For this reason, the instability
and its current sheet are also of interest for the coronal heating
problem. Following Parker's original suggestion,\cite{Parker1972}
many authors \cite{GalsgaardN1996,LongbottomRCS1998,NgB1998,MikicSV1989}
have shown that random shuffling of the coronal magnetic fieldlines
by photospheric motions progressively increases the current density,
resulting in sporadic energy release. How these spatially intermittent
currents are produced, and in particular whether MHD instability plays
any role, remains unclear. The kink instability is an interesting
possible model of energy buildup and release. Line-tied kinking has
also been captured in laboratory experiments; a recent example was
reported in Ref. \cite{BergersonFFHKSS2006}.

However, the theory of the kink mode developed for periodic plasma
does not carry over to line-tied systems such as coronal loops in
a straightforward manner. It is generally agreed that line-tying is
stabilizing, in the sense that a system with periodic boundary conditions
is more unstable than an identical system of the same length with
line-tied boundary conditions. Stabilization arises from the extra
tension force associated with the anchored footpoints. Roughly, the
system should be stable if the Alfv\'en travel time along the cylinder
is less than the inverse of the growth rate in a periodic system.
The first quantitative analysis of the effect of line-tying on stability
was given by Raadu,\cite{Raadu1972} who minimized the energy of a
restricted class of perturbations. However, there is a deeper issue.
Eigenfunctions with helical symmetry cannot satisfy boundary conditions
on surfaces of constant $z$. Therefore, the significance of rational
surfaces in aperiodic systems is open to question. Nevertheless, numerical
studies \cite{LionelloSEV1998,LionelloVEM1998,GerrardH2003,GerrardH2004,GerrardHB2004}
have shown that when a line-tied flux tube is twisted by motions of
its endpoints, it eventually kinks, and the kinked state has a thin
current layer (whether it is a current singularity or not is very
difficult to examine by a numerical calculation). On the other hand,
it has been shown quite rigorously that current sheets with the simple
ribbon topology found from kink theory cannot appear in line-tied
magnetic fields as long as their endpoint motion is a smooth function
of position.\cite{vanBallegooijen1985,CowleyLS1997} Thus, the question
of what sets the current density in a kinked, line tied system is
still open to investigation.

The purpose of ths paper is to clarify the relationship between the
periodic and line-tied kink instabilities. In particular, we address
the formation of thin current layers in line tied systems. In a periodic
system, the rational surfaces constitute a natural set of separatrices,
where current sheets can form. In a line-tied configuration, the notion
of rational surfaces is absent; as such, it is not clear a priori
where a thin current layer or current sheet, if any, would form. The
present study addresses this important issue. And the answer, in short,
is that the steep gradient in a line-tied system will appear at the
resonant surface corresponding to the fastest growing periodic mode
in a infinitely long system. A detailed comparison between the periodic
eigenmode and the line-tied eigenmode is made. Through the comparison,
the effects of line-tying on the growth rate and the eigenfunction
can be clarified.

We limit ourselves to the linearized problem in this study, and leave
the nonlinear problem to future work. This paper is organized as follows.
Sec. \ref{sec:Model-and-Eigenmode} give the details of the model
and the governing linearized equations. Sec. \ref{sec:Numerical-Solution-of}
presents the main results: comparsion of the periodic eigenmode and
the line-tied one. We mostly concentrate on the fastest growing mode
in both cases. Details are given for how the latter approach the former
in the limit when the system length $L$ goes to infinity. An important
quantity --- the thickness of the internal layer --- is found to follow
a scaling law in that limiting process. Some other scaling laws regarding
the critical length, the internal layer thickness and the mode localization
along the axial direction at marginality are also found. We conclude
and give a discussion in Sec. \ref{sec:Discussion}. Two appendices
follow the main text. Appendix \ref{sec:Numerical-Method} gives the
details of the numerical methods, and a semi-analytic calculation
for the critical length and the growth rate is detailed in Appendix
\ref{sec:semi-analytic}.

\section{Model and Eigenmode Equations\label{sec:Model-and-Eigenmode}}

We assume that the plasma pressure is negligible, $p=0$, and that
the equilibrium plasma density $\rho=\mbox{const}$, for simplicity.
In cylindrical coordinates $(r,\phi,z)$, the equilibrium magnetic
field is \begin{equation}
\mathbf{B}=B_{\phi}(r)\bm{\hat{\phi}}+B_{z}(r)\mathbf{\hat{z}},\label{eq:B-field}\end{equation}
which satisfies the force balance equation\begin{equation}
-\frac{d}{dr}\left(\frac{B^{2}}{8\pi}\right)-\frac{B_{\phi}^{2}}{4\pi r}=0.\label{eq:equilibrium}\end{equation}
Assuming $e^{\gamma t}$ time dependence, the linearized ideal MHD
equation can be expressed in terms of the Lagrangian displacement
$\bm{\xi}$ as \begin{equation}
\gamma^{2}\rho\bm{\xi}=\frac{1}{4\pi}(\nabla\times(\nabla\times(\bm{\xi}\times\mathbf{B})))\times\mathbf{B}+\frac{\nabla\times\mathbf{B}}{4\pi}\times(\nabla\times(\bm{\xi}\times\mathbf{B}))=0.\label{eq:linearizedmhd}\end{equation}
It is convenient to decompose the displacement $\bm{\xi}$ into the
radial, perpendicular, and parallel components as \begin{equation}
\bm{\xi}=\xi_{r}\mathbf{\hat{r}}+\xi_{\eta}\bm{\hat{\eta}}+\xi_{\parallel}\mathbf{\hat{b}},\label{eq:decompose}\end{equation}
where \begin{equation}
\bm{\hat{\eta}}=(B_{z}\bm{\hat{\phi}}-B_{\phi}\mathbf{\hat{z}})/B,\label{eq:eta}\end{equation}
and $\mathbf{\hat{b}}$ is the unit vector along the equilibrium magnetic
field. 

In the equilibrium, the $\phi$ direction is ignorable; we assume
azimuthal dependence $e^{im\phi}$. Taking the $\mathbf{\hat{b}}$
component of Eq. (\ref{eq:linearizedmhd}) gives $\xi_{\parallel}=0$.
The remaining two independent components of (\ref{eq:linearizedmhd})
are:

\begin{eqnarray}
\gamma^{2}\xi_{r} & = & V_{A}^{2}\xi_{r}''+\frac{V_{Az}^{2}-V_{A\phi}^{2}}{r}\xi_{r}'-\left(\frac{2V_{A\phi}V_{A\phi}'}{r}+\frac{V_{A}^{2}}{r^{2}}\right)\xi_{r}+\left(V_{Az}\partial_{z}+\frac{imV_{A\phi}}{r}\right)^{2}\xi_{r}\nonumber \\
 &  & -V_{A\phi}V_{A}\partial_{z}\xi_{\eta}'-\left(\frac{2V_{A\phi}V_{A}}{r}+(V_{A}V_{A\phi})'\right)\partial_{z}\xi_{\eta}+i\frac{mV_{A}V_{Az}}{r}\xi_{\eta}'\nonumber \\
 &  & +i\frac{m}{r}\left((V_{A}V_{Az})'-\frac{V_{A}V_{Az}}{r}\right)\xi_{\eta},\label{eq:eigen1}\end{eqnarray}
\begin{equation}
\gamma^{2}\xi_{\eta}=-V_{A}V_{A\phi}\partial_{z}\left(\xi_{r}'-\frac{\xi_{r}}{r}\right)+i\frac{mV_{A}V_{Az}}{r}\left(\xi_{r}'+\frac{\xi_{r}}{r}\right)+iV_{A}^{2}\left(\partial_{z}^{2}-\frac{m^{2}}{r^{2}}\right)\xi_{\eta},\label{eq:eigen2}\end{equation}
where primes denote $\partial_{r}$;$V_{Az}=B_{z}/\sqrt{4\pi\rho}$,
$V_{A\phi}=B_{\phi}/\sqrt{4\pi\rho}$ , and $V_{A}=B/\sqrt{4\pi\rho}$,
respectively.

In this work we are mostly interested in systems with a strong guide
field, i.e., $V_{Az}\gg V_{A\phi}$. In that case, the kink mode is
nearly incompressible and the growth rate is much smaller than the
Alfv\'en frequency, i.e., $\gamma^{2}\ll V_{A}^{2}/a^{2}$, where
$a$ is a characteristic perpendicular length scale of the flux tube.
That means that large terms ($\sim O(V_{A}^{2}/a^{2})\left|\bm{\xi}\right|$)
on the right-hand side (RHS) of Eqs. (\ref{eq:eigen1}) and (\ref{eq:eigen2})
nearly cancel each other; the small unbalanced remainders are balanced
by the inertia term on the left-hand side (LHS). The cancellation
of large terms could lead to a significant loss in accuracy when the
eigenmode equations are solved numerically. To avoid that we reformulate
the eigenmode equations as follows. Let \begin{equation}
\xi_{\eta}=i\frac{rV_{Az}}{mV_{A}}\left(\xi_{r}'+\frac{\xi_{r}}{r}\right)+\tilde{\xi_{\eta}},\label{eq:eta-decompose}\end{equation}
where the first part on the RHS gives (approximate)
incompressibility and $\tilde{\xi_{\eta}}$ is a small correction.
Substituting (\ref{eq:eta-decompose}) into the eigenmode equations,
we have \begin{eqnarray}
\gamma^{2}\xi_{r} & = & V_{A\phi}^{2}\xi_{r}''+\left(\frac{V_{A\phi}^{2}}{r}+2V_{A\phi}V_{A\phi}'\right)\xi_{r}'+\frac{(1-m^{2})V_{A\phi}^{2}}{r^{2}}\xi_{r}\nonumber \\
 &  & +\left(V_{Az}^{2}\partial_{z}^{2}+\left(\frac{2i(m^{2}-1)V_{A\phi}V_{Az}}{mr}+\frac{iV_{A\phi}^{3}}{mrV_{Az}}+\frac{i(V_{A\phi}^{2}-V_{Az}^{2})V_{A\phi}'}{mV_{Az}}\right)\partial_{z}\right)\xi_{r}\nonumber \\
 &  & -\frac{irV_{Az}V_{A\phi}}{m}\partial_{z}\xi_{r}''+\left(\frac{iV_{A\phi}^{3}}{mV_{Az}}+\frac{ir(V_{A\phi}^{2}-V_{Az}^{2})V_{A\phi}'}{mV_{Az}}-\frac{4iV_{A\phi}V_{Az}}{m}\right)\partial_{z}\xi_{r}'\nonumber \\
 &  & -V_{A\phi}V_{A}\partial_{z}\tilde{\xi_{\eta}}'-\left(\frac{2V_{A\phi}V_{A}}{r}-\frac{V_{A\phi}^{3}}{V_{A}r}+V_{A}V_{A\phi}'\right)\partial_{z}\tilde{\xi_{\eta}}+i\frac{mV_{A}V_{Az}}{r}\tilde{\xi_{\eta}}'\nonumber \\
 &  & +i\frac{m}{r}\left(-\left(\frac{V_{Az}}{V_{A}}+\frac{V_{A}}{V_{Az}}\right)\frac{V_{A\phi}^{2}}{r}-\frac{V_{A}V_{A\phi}V_{A\phi}'}{V_{Az}}-\frac{V_{A}V_{Az}}{r}\right)\tilde{\xi_{\eta}},\label{eq:reform1}\end{eqnarray}
\begin{eqnarray}
\gamma^{2}\left(-\frac{rV_{Az}}{mV_{A}}\left(\xi_{r}'+\frac{\xi_{r}}{r}\right)+i\tilde{\xi_{\eta}}\right) & = & -iV_{A}V_{A\phi}\partial_{z}\left(\xi_{r}'-\frac{\xi_{r}}{r}\right)-\frac{rV_{A}V_{Az}}{m}\partial_{z}^{2}\left(\xi_{r}'+\frac{\xi_{r}}{r}\right)\nonumber \\
 &  & +iV_{A}^{2}\left(\partial_{z}^{2}-\frac{m^{2}}{r^{2}}\right)\tilde{\xi_{\eta}}.\label{eq:reform2}\end{eqnarray}
In this way the large terms on the RHS have been cancelled explicitly;
all terms are $O(V_{A\phi}^{2}/a^{2})\left|\bm{\xi}\right|$ or smaller. 

To make the numerical solution easier, we assume the existence of
a conducting wall at $r=r_{0}$, which is made sufficiently far away
that stability is only weakly affected. Two conducting plates at $z=\pm L/2$
anchor the magnetic field footpoints. Under these assumptions, the
boundary conditions are $\xi_{r}|_{z=\pm L/2}=\tilde{\xi_{\eta}}|_{z=\pm L/2}=0$,
and $\xi_{r}|_{r=r_{0}}=0$. The regularity condition at $r=0$ requires
that $\xi_{r}\sim r^{|m|-1}$ and $\tilde{\xi_{\eta}}\sim r^{|m|+1}$
as $r$ approaches zero.

\section{Numerical Solution of Eigenmode Equations \label{sec:Numerical-Solution-of}}

In this section we present the numerical solutions of the eigenmode
equations. The numerical method is described in Appendix \ref{sec:Numerical-Method}.
The model system we solve in this work has the following force free
equilibrium profiles:\begin{equation}
V_{A\phi}=\epsilon r\exp(-r^{2}/2),\label{eq:Vaphi}\end{equation}
\begin{equation}
V_{Az}=\sqrt{1+\epsilon^{2}(1-r^{2})\exp(-r^{2})}.\label{eq:Vaz}\end{equation}
This simple system has only three free parameters: $\epsilon$, $L$,
and $r_{0}$. We fix the conducting outer boundary at $r_{0}=5$,
sufficiently far that it only weakly affects the stability and the
eigenfunction. The parameter $\epsilon\sim V_{A\phi}/V_{Az}$ is usually
a small parameter in tokamak and coronal loop applications; in this
work we systematically explore the region $\epsilon\le1$. For a given
equilibrium, we solve the eigenmode equations for different $L$.
We are particularly interested in studying how the growth rate and
the thickness of the internal layer vary with the system length $L$.

\subsection{Periodic Solutions}

Before going into the results for line-tied systems, we present the
results of the corresponding periodic systems. These results will
serve as references to the line-tied solutions.

The analytical theory of the $m=1$ ideal internal kink mode in the
limit $\epsilon\ll1$ is given by Rosenbluth, Dagazian, and Rutherford
in Ref. \cite{RosenbluthDR1973} (hereafter RDR). Key results relevant
to the present study are summarized as follows: (1) For an unstable
eigenmode with spatial dependence $\exp(ikz+im\phi)$, with $m=\pm1$,
the eigenfunction $\xi_{r}(r)$ has a steep gradient (the internal
layer) at the resonant surface $r=r_{s}$, i.e., where $\mathbf{k}\cdot\mathbf{B}=kB_{z}+mB_{\phi}/r_{s}=0$.
Outside of the internal layer, $\xi_{r}\simeq\xi_{a}=\mbox{const}$
for $r<r_{s}$ and $\xi_{r}\simeq0$ for $r>r_{s}$. (2) The growth
rate of the unstable mode is approximately (there is a misprint of
a factor of two in the original paper)\begin{equation}
\gamma\simeq-\frac{\pi}{\left|(\mathbf{k}\cdot\mathbf{V}_{A})'\right|_{r_{s}}r_{s}^{3}}\int_{0}^{r_{s}}g_{1}dr,\label{eq:gamma-RDR}\end{equation}
where \begin{equation}
g_{1}=k^{2}r(3k^{2}r^{2}V_{Az}^{2}+2kmrV_{Az}V_{A\phi}-V_{A\phi}^{2}).\label{eq:g1}\end{equation}
(3) The solution within the internal layer is approximately\begin{equation}
\xi_{r}\simeq\frac{1}{2}\xi_{a}\left(1-\frac{2}{\pi}\tan^{-1}\left(\frac{x\left|(\mathbf{k}\cdot\mathbf{V}_{A})'\right|_{s}}{\gamma}\right)\right),\label{eq:internal-layer-RDR}\end{equation}
where $x\equiv r-r_{s}$.

Some scaling laws can be deduced from these results. First, we have
$\gamma\propto\epsilon^{3}$ in the following sense. In the limit
$\epsilon\ll1$, $V_{Az}$ is approximately constant. If we vary $\epsilon$
and let $k\propto\epsilon$, the resonant surface will be approximately
located at a fixed radius, and the growth rate $\gamma$ will be proportional
to $\epsilon^{3}$, since $g_{1}\propto\epsilon^{4}$ and $\left|(\mathbf{k}\cdot\mathbf{V}_{A})'\right|_{r_{s}}\propto\epsilon$.
Likewise, the thickness of the internal layer is proportional to $\epsilon^{2}$.
To be precise, we define the thickness of the internal layer $\Delta$
as the distance between the two radii where $\xi_{r}=(3/4)\xi_{a}$
and where $\xi_{r}=(1/4)\xi_{a}$. Eq. (\ref{eq:internal-layer-RDR})
gives \begin{equation}
\Delta\simeq2\gamma/\left|(\mathbf{k}\cdot\mathbf{V_{A}})'\right|_{r_{s}}.\label{eq:thickness_RDR}\end{equation}
Hereafter we will use the subscript {}``$0$'' to denote properties
of or related to the fastest growing periodic mode: for a given equilibrium,
the fastest growing mode appears at the wavenumber $k=k_{0}$, with
the growth rate $\gamma=\gamma_{0}$, and the internal layer thickness
$\Delta_{0}$. From the scaling laws we have $k_{0}\propto\epsilon$
, $\gamma_{0}\propto\epsilon^{3}$, and $\Delta_{0}\propto\epsilon^{2}$.

\begin{figure}[t]
\includegraphics[scale=0.4]{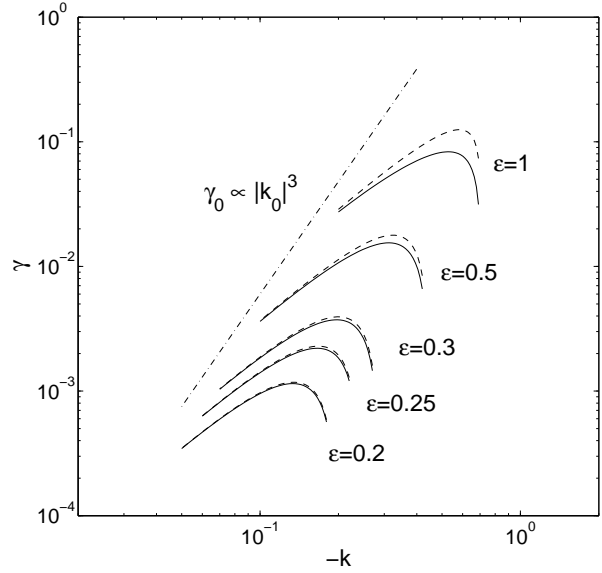}

\caption{The growth rate $\gamma$ as a function of $k$, for different $\epsilon$.
Solid lines are the growth rates calculated by the code, and dashed
lines are the approximate growth rates calculated from Eq. (\ref{eq:gamma-RDR}).
As expected, the agreement becomes better for smaller $\epsilon$.
The dashdot line indicates the scaling law, $\gamma_{0}\propto\left|k_{0}\right|^{3}$.\label{cap:gamma-vs-k}}
\end{figure}

\begin{figure}[t]
\includegraphics[scale=0.8]{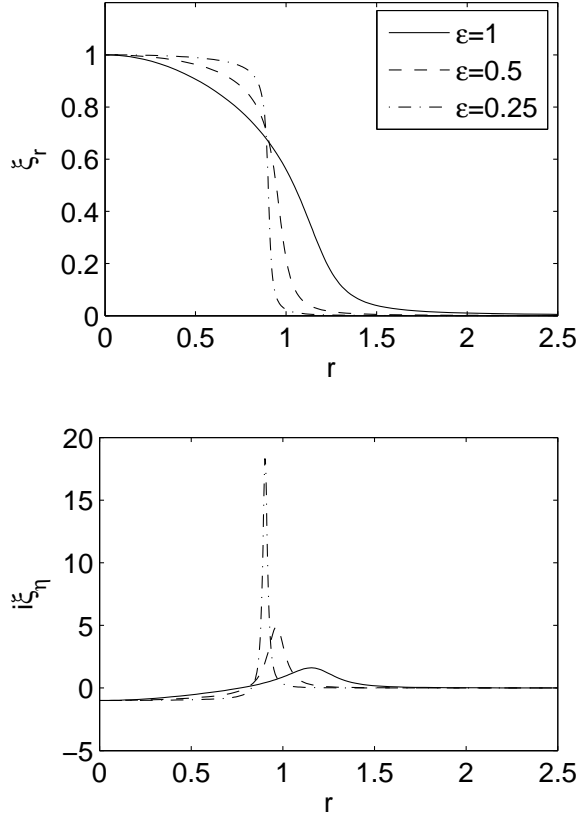}

\caption{Periodic eigenfunctions of the fastest growing modes for $\epsilon=1$,
$0.5$, $0.25$. The radial displacement $\xi_{r}$ has a jump at
$r\simeq1$. The jump becomes steeper, and the twist ($\xi_{\eta}$)
becomes more localized, for smaller $\epsilon$. Note that only the
solutions within $0\le r\le2.5$ are shown; outside of that the solutions
are vanishingly small. The eigenfunctions are normalized such that
$\xi_{r}(0)=1$. \label{cap:periodic-eigen}}
\end{figure}

Figure \ref{cap:gamma-vs-k} shows the growth rate $\gamma$ as a
function of $k$, for different $\epsilon$. Solid lines denote the
growth rate calculated by the code, and dashed lines are the approximate
growth rate calculated from Eq. (\ref{eq:gamma-RDR}). As expected,
the agreement becomes better for smaller $\epsilon$. For each $\epsilon$,
the growth rate peaks at $k=k_{0}$, and the corresponding $\gamma_{0}$
follows the scaling law, $\gamma_{0}\propto\left|k_{0}\right|^{3}$,
as indicated by the dashdot line. The numerical values of $k_{0}$,
$\gamma_{0}$ , and $\Delta_{0}$ for different $\epsilon$ are summarized
in Table \ref{cap:Summary-table}. Figure \ref{cap:periodic-eigen}
shows the eigenfunctions of the fastest growing modes for $\epsilon=1,$
$0.5$, $0.25$. The radial displacement $\xi_{r}$ of each shows
a jump at $r\simeq1$. The jump becomes steeper, and the twist ($\xi_{\eta}$)
becomes more localized, for smaller $\epsilon$.

Notice that here the axial wavenumber $k$ is treated as a continuous
parameter. This treatment actually corresponds to periodic eigenmodes
in an \emph{infinitely} long system. For a periodic system with a
finite length $L$, the wavenumber $k$ is quantized as $k=2n\pi/L$,
with integer $n$. As such, only a finite number of unstable modes
are present in a finite length system.

\subsection{Line-tied Solutions}

We start by observing some general characteristics of the eigenmodes
in line-tied systems, then move on to the scaling laws obtained from
analyzing the numerical solutions. We focus only on the fastest growing
mode for a given configuration in the first two sections. Higher harmonics
are briefly discussed in section \ref{sub:Higher-Harmonics}.

\subsubsection{General Observations\label{sub:General-Observations}}

\begin{figure}[t]
\includegraphics[scale=0.45]{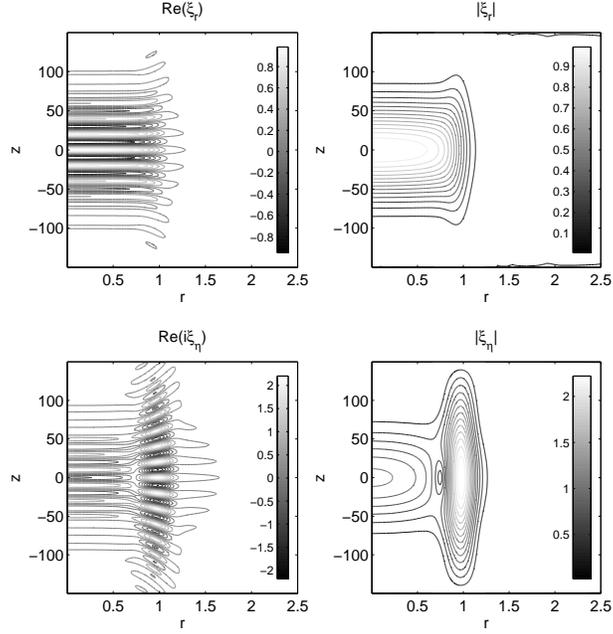}

\caption{The fastest growing mode for the case $\epsilon=0.5$, $L=300$.
The two panels on the left show the real part of $\xi_{r}$ and $i\xi_{\eta}$.
The imaginary parts are not shown; they are similar to the real parts,
only with different phases. The two panels on the right show the module
of each component. The eigenfunctions are normalized such that $\xi_{r}|_{r=z=0}=1$.
\label{cap:e05L300eig}}
\end{figure}

Figure \ref{cap:e05L300eig} shows the eigenfunctions of the fastest
growing mode for $\epsilon=0.5$, $L=300$, which demonstrate some
general characteristics observed throughout all the cases we have
tried. The first thing to notice is that there are many oscillations
along the $z$ direction. The wavelength is approximately the same
as the corresponding fastest growing mode in the periodic case. Second,
the radial displacement also has a jump at $r\simeq1$, the same location
as the jump of the fastest growing periodic mode. Third, the eigenmode
is more or less localized to the center, rather than being broadened
out to the whole $z$ domain, with an envelope $\sim\cos(\pi z/L)$,
which seems to minimize the field line bending due to line-tying.
In fact, the $\cos(\pi z/L)$ envelope is what Raadu used in his energy
principle analysis,\cite{Raadu1972} yet the numerical solutions suggest
otherwise. We will come back to this issue later.

\begin{figure}[t]
\includegraphics[scale=0.45]{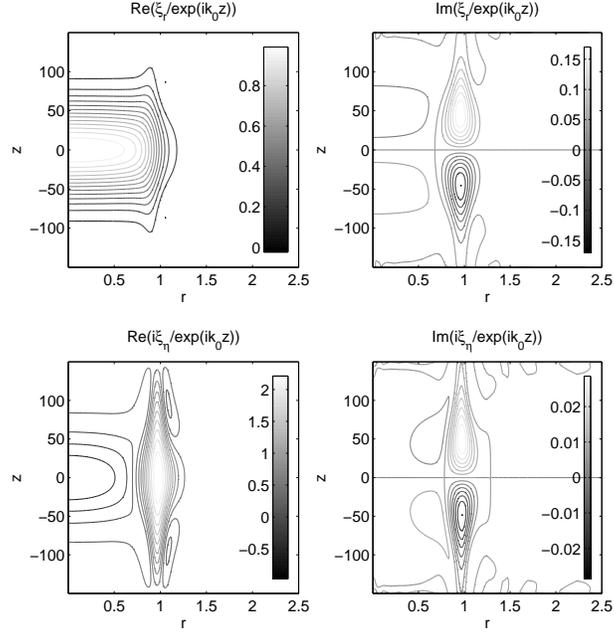}

\caption{The fastest growing mode for $\epsilon=0.5$, $L=300$, divided by
$\exp(ik_{0}z)$ to filter out the fast oscillations. The remaining
{}``envelopes'' are slowly varying functions along the $z$ direction.
\label{cap:e05L300eig-filtered}}
\end{figure}

Since the wavelength along $z$ is approximately the same as that
of the fastest growing periodic mode, one may {}``filter out'' the
fast oscillations along $z$ by dividing the solutions by $\exp(ik_{0}z)$.
The results are shown in Figure \ref{cap:e05L300eig-filtered}. The
remaining {}``envelopes'' of the eigenfunctions become slowly varying
along the $z$ direction. This feature has been utilized into the
choice of the basis functions of the numerical method, detailed in
Appendix \ref{sec:Numerical-Method}. In short, instead of having
to resolve the fast oscillations along $z$, we only need to resolve
the slowly varying envelopes, therefore many fewer basis functions
are needed. 

\begin{figure}[t]
\includegraphics[scale=0.45]{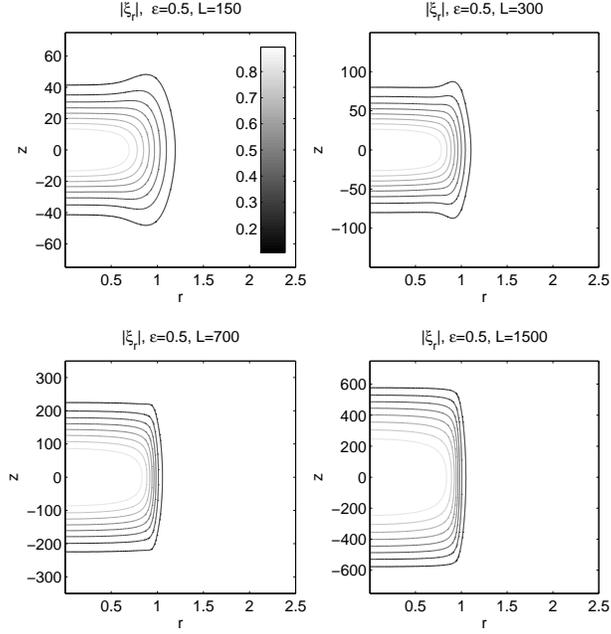}

\caption{Contour plots of $\left|\xi_{r}\right|$for $\epsilon=0.5$, $L=150$,
$300$, $700$, and $1500$. All solutions are normalized to $\xi_{r}|_{r=z=0}=1$.
The contours correspond to $\left|\xi_{r}\right|=0.1$, 0.2, ...,
0.8. As the system length becomes longer, the jump in $\left|\xi_{r}\right|$
becomes narrower, and the eigenfunction becomes broader along the
$z$ direction. For a really long system, the $\cos(\pi z/L)$ envelope
becomes a good approximation.\label{cap:e05eig-Lscan}}
\end{figure}

\begin{figure}[t]
\includegraphics[scale=0.7]{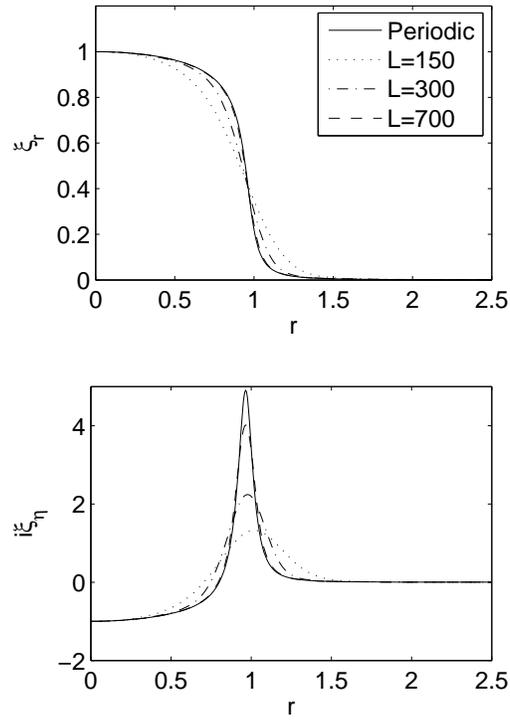}

\caption{Eigenfunctions at the midplane, $z=0$, for different $L$, as compared
to the fastest growing periodic eigenfunctions. $\epsilon=0.5$.\label{cap:periodic-vs-linetied}}
\end{figure}

\begin{figure}[t]
\includegraphics[scale=0.6]{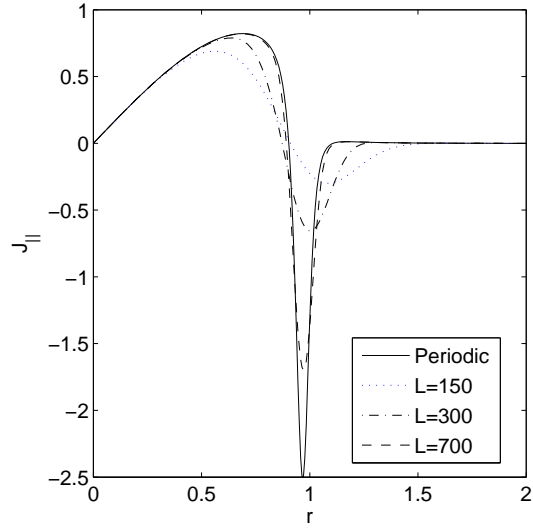}

\caption{The perturbed parallel current at the midplane $z=0$. The thin current
layer of the periodic case is smoothed in line-tied cases. The line-tied
case approaches the periodic one as $L$ becomes larger.\label{cap:The-perturbed-parallel}}
\end{figure}

Figure \ref{cap:e05eig-Lscan} shows the fastest growing mode for
different $L$, with a fixed $\epsilon=0.5$. We observe that as $L$
becomes larger, the jump of the radial displacement becomes steeper,
and the eigenfunction becomes broader along $z$. For a really long
$L$, the envelope $\cos(\pi z/L)$ becomes a good approximation.
Figure \ref{cap:periodic-vs-linetied} shows the eigenfunctions at
$z=0$, for various $L$, as compared to the periodic fastest growing
mode. As $L$ becomes larger, the line-tied eigenfunctions at the
midplane approach the periodic ones. For the case $L=1500$, which
is not shown on the plot, the eigenfunctions at the midplane are virtually
indistinguishable from the periodic ones. The parallel component of
the perturbed current $J_{\parallel}\equiv\mathbf{\hat{b}}\cdot\nabla\times\nabla\times(\bm{\xi}\times\mathbf{V}_{A})$
at the midplane is shown in Figure \ref{cap:The-perturbed-parallel}.
We see the thin current layer of the periodic solution is smoothed
in line-tied cases. As the system length becomes longer, the line-tied
solution approaches the periodic one. It should be pointed out, however,
that the aspect ratio of these systems are much larger than in natural
systems such as coronal loops.

\begin{figure}[t]
\includegraphics[scale=0.5]{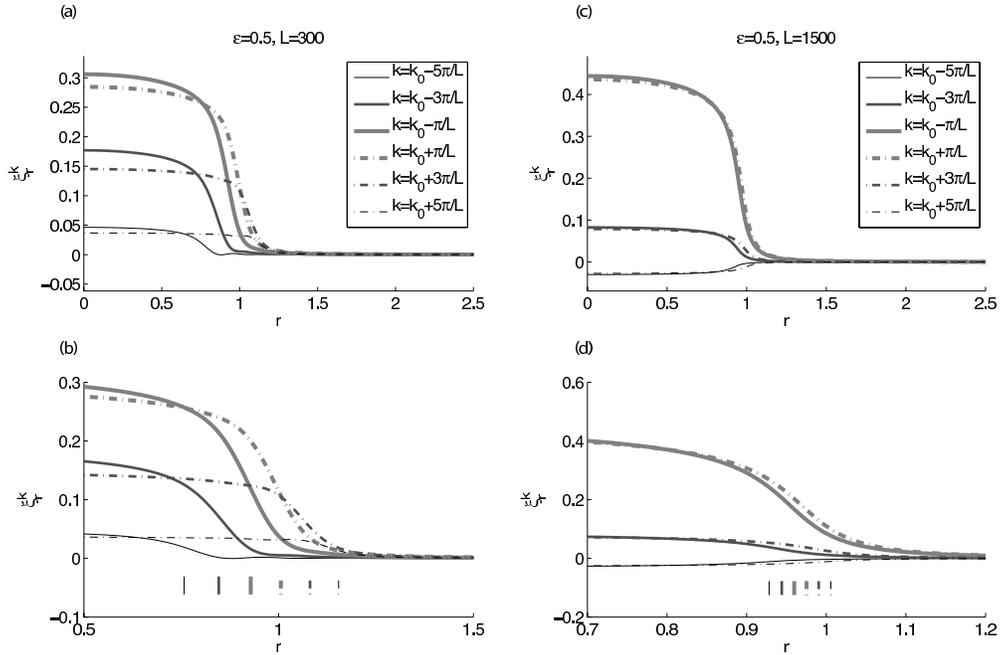}

\caption{(a) The six most significant {}``shifted'' Fourier components for
the case $\epsilon=0.5$, $L=300$. (b) Expanded view of the jump
at $r\simeq1$. Vertical lines indicate the $\mathbf{k}\cdot\mathbf{B}=0$
surfaces corresponding to each $k$. (c) and (d) are the same as (a)
and (b), for a longer system with $L=1500$. \label{cap:e05fr}}
\end{figure}

Additional insight may be obtained by decomposing the eigenfunctions
into Fourier harmonics. Because the wavelength along $z$ is approximately
the same as the wavelength corresponds to $k_{0}$, we consider the
following {}``shifted'' Fourier decomposition:\begin{equation}
\xi_{r}=\sum_{n=-\infty}^{\infty}\xi_{r}^{n}\exp\left(i\left(k_{0}+(2n-1)\pi/L\right)z\right).\label{eq:fourier}\end{equation}
Figure \ref{cap:e05fr} (a) shows the six most significant Fourier
components for the case $\epsilon=0.5$, $L=300$. We observe that
they roughly form three pairs. The two components in each pair have
approximately the same amplitude such that they nearly cancel each
other at the ends, $z=\pm L/2$. However, the expanded view (b) about
the jump at $r\simeq1$ shows that each component has a jump at its
own $\mathbf{k}\cdot\mathbf{B}=0$ surface, precluding cancellation
within the narrow internal layer. As a result, many (actually an infinite
number of) Fourier harmonics are needed to achieve a full cancellation
at the ends. For a longer system, the distance between neighboring
$\mathbf{k}\cdot\mathbf{B}=0$ surfaces becomes smaller. As shown
in Figure \ref{cap:e05fr} (c)(d), for the case $\epsilon=0.5$, $L=1500$,
the two dominant components are nearly identical throughout the whole
domain, and the components of other harmonics are very small. One
may anticipate that in the limit $L\rightarrow\infty$ the solution
is completely dominated by two components, which give a $z$ dependence
$\sim\exp(ik_{0}z)\cos(\pi z/L)$. This is similar to the external
kink mode solution calculated by Hegna \cite{Hegna2004} and Ryutov
et. al. \cite{RyutovCP2004}

\subsubsection{Scaling Laws and {}``Universalities''\label{sub:Scaling-Laws}}

It would be desirable if we could deduce some scaling laws from the
numerical solutions. In particular, we are interested in how the internal
layer thickness, the growth rate, the mode localization along $z$,
the critical length for instability, etc., vary with $\epsilon$ or
$L$. To be precise, let us define the quantities to be measured.
The thickness $\Delta$ of the internal layer is measured at $z=0$.
Let $\xi_{a}\equiv\left|\xi_{r}\right|_{r=0,z=0}$. The thickness
$\Delta$ is defined as the distance between the radii where $\left|\xi_{r}\right|=(3/4)\xi_{a}$
and $\left|\xi_{r}\right|=(1/4)\xi_{a}$. The length $l$ of an eigenmode
along $z$ is measured at $r=0$, and defined as the distance between
the two points where $\left|\xi_{r}\right|=(1/2)\xi_{a}$. We use
the subscript {}``$c"$ to denote properties associated with marginal
stability: $L_{c}$ is the critical length for a given $\epsilon$;
$l_{c}$ is the length, and $\Delta_{c}$ is the internal layer thickness,
of the marginally stable eigenmode. 

Table \ref{cap:marginal-scaling} summarizes $L_{c}$, $\Delta_{c}$,
and $l_{c}$ for various $\epsilon$. For small $\epsilon$, we observe
the following scaling laws: $L_{c}\propto1/\epsilon^{3}$, $(\Delta_{c}-\Delta_{0})/\Delta_{0}\propto1/\epsilon$,
and $l_{c}/L_{c}\propto\epsilon$. The scaling $L_{c}\propto1/\epsilon^{3}$
can be understood as follows. From a semi-analytic calculation (see
Appendix \ref{sec:semi-analytic} for details), we have an estimate
for the critical length\begin{equation}
L_{c}\simeq2\left|V_{Az}\right|_{r_{s}}/\gamma_{0}\label{eq:estimate}\end{equation}
where $V_{Az}$ is evaluated at the resonant surface of the fastest
growing periodic mode. For $\epsilon=$1, 0.5, 0.3, 0.25, 0.2, 0.15
this estimate gives $L_{c}\simeq$22.9, 130, 538, 910, 1746, 4079.
Compared with the $L_{c}$ summarized in Table \ref{cap:marginal-scaling},
the agreement is very good for small $\epsilon$. Since $V_{Az}\simeq\text{const}$
and $\gamma_{0}\propto\epsilon^{3}$ for small $\epsilon$, the scaling
law follows. In the astrophysics literature, more often the critical
twist of the magnetic field for a given flux tube length is considered.
Let us consider the twist at the critical length:\begin{equation}
\Delta\phi_{c}\propto\frac{L_{c}B_{\phi}}{rB_{z}}\propto\frac{1}{\epsilon^{2}}\propto L_{c}^{2/3}.\label{eq:twist}\end{equation}
If we consider the flux tube length $L$ as given, and twist up the
magnetic field to make it unstable, the critical twist scales as $L^{2/3}$. 

\begin{figure}[t]
\includegraphics[scale=0.45]{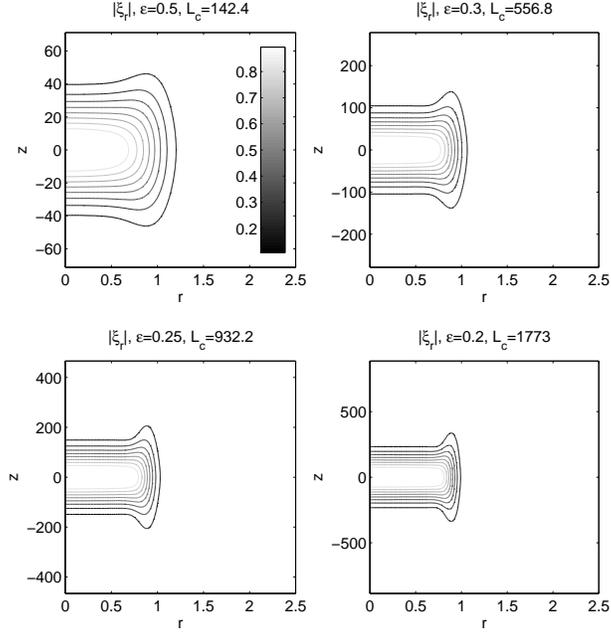}

\caption{$\left|\xi_{r}\right|$ of the marginally stable eigenmodes for $\epsilon=0.5$,
$0.3$, $0.25$, $0.2$. The eigenfunctions are normalized to $\xi_{r}|_{r=0,z=0}=1$.
The contours correspond to $\left|\xi_{r}\right|=$0.1, 0.2, ...,
0.8.\label{cap:marginal}}
\end{figure}

The quantity $(\Delta_{c}-\Delta_{0})/\Delta_{0}$ measures the relative
increase in the internal layer thickness due to line-tying, and $l_{c}/L_{c}$
measures the localization along $z$ of the marginal mode. The scaling
law $(\Delta_{c}-\Delta_{0})/\Delta_{0}\propto1/\epsilon$ indicates
that the increase in internal layer thickness is greater for smaller
$\epsilon$, at marginal stability. And the scaling law $l_{c}/L_{c}\propto\epsilon$
indicates that the marginally stable mode becomes more and more localized
along $z$ for smaller $\epsilon$. Figure \ref{cap:marginal} shows
$\left|\xi_{r}\right|$ of the marginally stable eigenmodes for various
$\epsilon$. The localization along $z$ is evident for smaller $\epsilon$.
Note that the trend of localization along $z$ and broadening of the
internal layer are reciprocal to each other. This is consistent with
the understanding based on Fourier mode decomposition --- localization
along $z$ means a lot of Fourier harmonics are involved, that also
means significant broadening of the internal layer. 

\begin{figure}[t]
\includegraphics[scale=0.6]{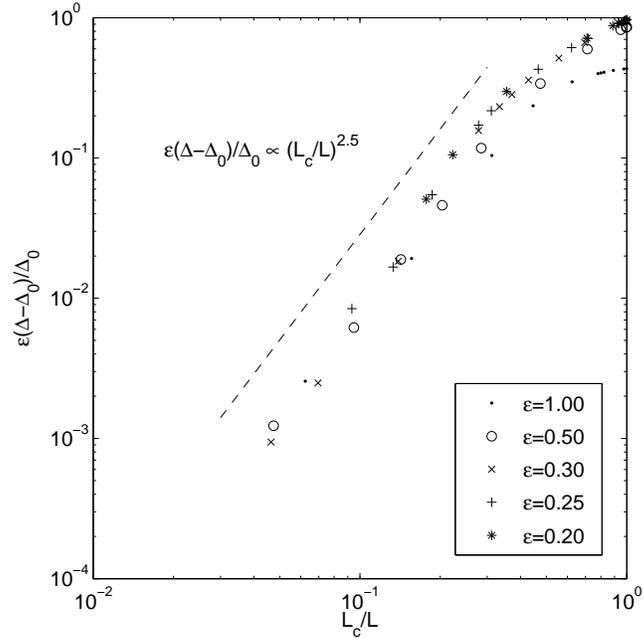}

\caption{$\epsilon(\Delta-\Delta_{0})/\Delta_{0}$ as a function of $L_{c}/L$,
for various $\epsilon$, in $\log-\log$ scale.\label{cap:delta-scaling}}
\end{figure}

Let us now look at the internal layer thickness $\Delta$ as a function
of $L$. We have already observed that $\Delta\rightarrow\Delta_{0}$
as $L\rightarrow\infty$. In an attempt to reveal some {}``universality'',
we plot the functional dependence for various $\epsilon$, with a
proper choice of variables and normalizations, on the same diagram.
A natural choice of normalization for $L$ is with respect to $L_{c}$;
therefore we choose $L_{c}/L$ as the horizontal axis. The choice
for the vertical axis is not all that obvious. Since $(\Delta_{c}-\Delta_{0})/\Delta_{0}\propto1/\epsilon$,
we choose $\epsilon(\Delta_{c}-\Delta_{0})/\Delta_{0}$ as the vertical
axis. With this choice of variables, the marginally stable solutions
for different $\epsilon$ all appear roughly at the same point. Figure
\ref{cap:delta-scaling} shows the resulting plot in $\log-\log$
scale. We observe that all curves for different $\epsilon$ roughly
coincide, and follow the scaling law $\epsilon(\Delta-\Delta_{0})/\Delta_{0}\propto(L_{c}/L)^{2.5}$.
The curves deviate from the scaling law near marginality. 

\begin{figure}[t]
\includegraphics[scale=0.9]{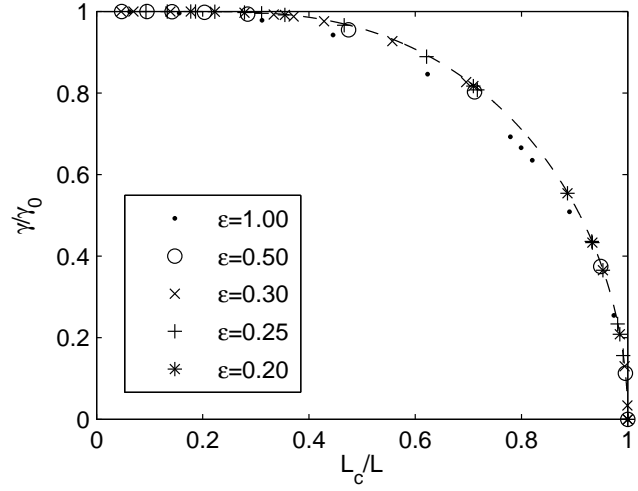}

\caption{$\gamma/\gamma_{0}$ as a function of $L_{c}/L$, for various $\epsilon$.
The dashed curve is analytic relation, Eq. (\ref{eq:gamma-L-universality}).
\label{cap:gamma-scaling}}
\end{figure}

Finally, let us look at how the growth rate varies with $L$. It is
found that $\gamma\rightarrow\gamma_{0}$ as $L\rightarrow\infty$.
We again try to plot the relationship between $\gamma$ and $L$ through
a proper choice of variables. A natural choice is $L_{c}/L$ as the
horizontal axis, and $\gamma/\gamma_{0}$ as the vertical axis. This
choice brings all marginally stable modes to the same point $(1,0)$
on the plot. Figure \ref{cap:gamma-scaling} shows the resulting plot,
for various $\epsilon$. Once again, all curves roughly coincide.
The relation between $\gamma/\gamma_{0}$ and $L_{c}/L$ can be obtained
by the semi-analytic calculation detailed in Appendix \ref{sec:semi-analytic}
as\begin{equation}
\frac{L_{c}}{L}=\frac{\gamma/\gamma_{0}}{\tanh^{-1}(\gamma/\gamma_{0})},\label{eq:gamma-L-universality}\end{equation}
which is in a good agreement with the numerical result. The line-tied
growth rate $\gamma$ approaches the fastest periodic growth rate
$\gamma_{0}$ rather quickly after the critical length is exceeded.
For $L=2L_{c}$, about $95\%$ of the periodic growth rate is recovered.

\subsubsection{Higher Harmonics\label{sub:Higher-Harmonics}}

\begin{figure}[t]
\includegraphics[scale=0.45]{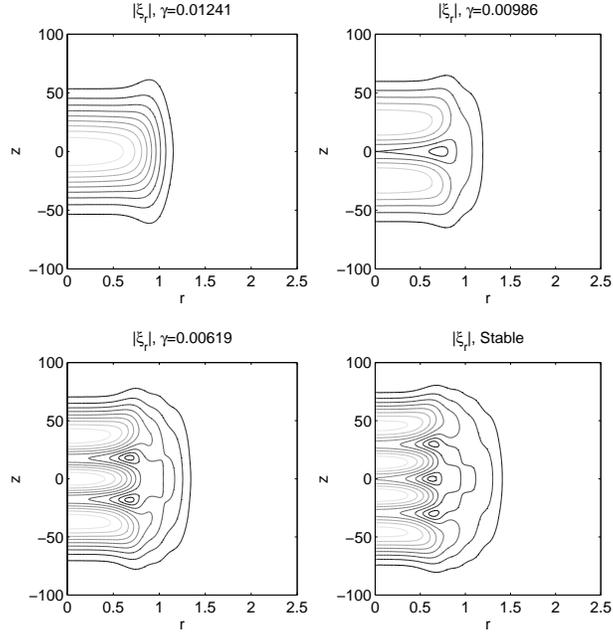}

\caption{$|\xi_{r}|$ of four harmonics for the case $\epsilon=0.5$, $L=200$.
Only three of them are unstable. The critical length is $L=142.4$.\label{cap:higher-harmonic-L200}}
\end{figure}

\begin{figure}[t]
\includegraphics[scale=0.45]{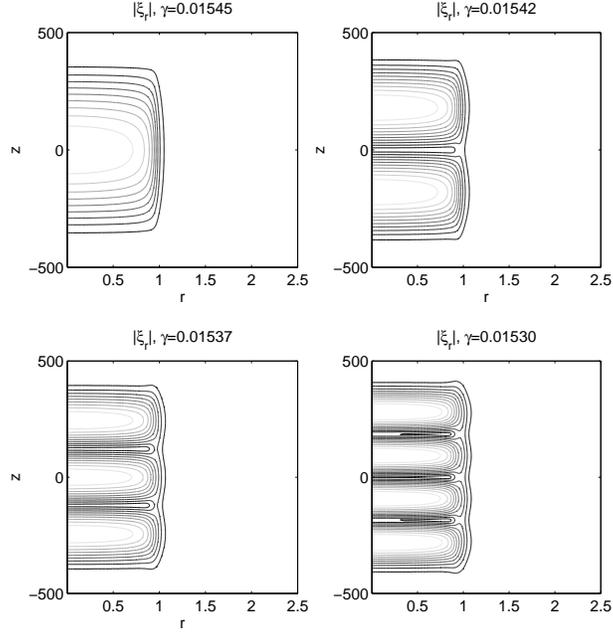}

\caption{$|\xi_{r}|$ of four unstable modes for the case $\epsilon=0.5$,
$L=1000$. \label{cap:higher-harmonic-L1000}}
\end{figure}

Thus far we have been focused on the fastest growing mode. Here we
briefly discuss the higher harmonics. Figure \ref{cap:higher-harmonic-L200}
shows four harmonics of the case $\epsilon=0.5$, $L=200$. Since
the length $L$ is not much longer than the critical length $L_{c}=142.4$,
only three unstable modes are present. The kinking central column
breaks into blobs for higher harmonics; the growth rate decreases
with the increase in the number of blobs (hence more field line bending).
Figure \ref{cap:higher-harmonic-L1000} shows the four most unstable
modes for the system with the same $\epsilon=0.5$, but much longer
$L=1000$. The higher harmonics still show the same general characteristics,
but the growth rates of them are very close to the fastest one. In
this case the physical significance of the fastest growing eigenmode
becomes questionable, since those higher harmonics are likely to play
an equally important role when the instability develops and reaches
nonlinear saturation.

\section{Summary and Discussion\label{sec:Discussion}}

In this paper we address the issue of the internal kink mode in a
line-tied system, as compared to that in a periodic system. We find
that the fastest growing internal kink mode in a line-tied system
possesses a steep internal layer, as does its periodic counterpart.
The internal layer locates at the resonant surface of the fastest
growing periodic mode in an \emph{infinitely} long system. Therefore,
even though the notion of rational surfaces is absent in a line-tied
system, we still have a rule of thumb as to where the current sheet
would be. Line-tying decreases the growth rate and increases the internal
layer thickness; only in the limit $L$ goes to infinity does the
line-tied mode approach the periodic one. 

In the small $\epsilon$ limit, the critical length can be estimated
rather accurately using Eq. (\ref{eq:estimate}), and the dependence
of the growth rate on $L$ is accurately represented by Eq. (\ref{eq:gamma-L-universality}).
Dimensionally, the critical length agrees with the physical picture
that the system becomes unstable when the Alfv\'en travel time along
the cylinder is longer than the inverse of the growth rate in a periodic
system. A similar estimate can also be derived by applying the RDR
asymptotic analysis to Raadu's energy principle calculation, which
gives \begin{equation}
L_{c}\simeq\frac{\pi V_{A}|_{r_{s}}}{\gamma_{0}}.\label{eq:estimate-raadu}\end{equation}
Functionally this is similar to Eq. (\ref{eq:estimate}), but significantly
overestimates the critical length by a factor of $\pi/2$. The reason
is that Raadu assumed a $\cos(\pi z/L)$ envelope in his trial function,
which is not a good approximation for the marginally stable mode,
as the latter is highly localized to the center.

Several interesting scaling laws are deduced from the numerical solutions,
most notably that $L_{c}\propto1/\epsilon^{3}$, $l_{c}\propto1/\epsilon^{2}$,
$(\Delta_{c}-\Delta_{0})/\Delta_{0}\propto1/\epsilon$, and $\epsilon(\Delta-\Delta_{0})/\Delta_{0}\propto(L_{c}/L)^{2.5}$.
However, the computational overhead increases rather rapidly as one
decreases $\epsilon$, due to the stringent scalings $\gamma_{0}\propto\epsilon^{3}$
and $\Delta_{0}\propto\epsilon^{2}$. As a result, the attainable
parameter range of the numerical solution is limited. It would be
desirable if the problem could be solved asymptotically in the limit
$\epsilon\ll1$, thereby the scaling laws can be {}``explained''.
Our attempt at asymptotic solution only succeeds partially. We hope
that the numerically found scaling laws can give some hint as to how
the full analysis could be carried out.

As for application to realistic situations, Eqs. (\ref{eq:estimate}),
(\ref{eq:gamma-L-universality}), and the observed scaling laws are
only valid in the small $\epsilon$ limit. For small $\epsilon$,
the critical aspect ratio $L/a$ can easily go up to hundreds or thousands
in order to make the flux tube unstable, and one may have trouble
finding such a long flux tube in nature. Furthermore, for a naturally
arising flux tube twisted up by footpoint motions, one can anticipate
that the tube length will not be much longer than the critical length
(or equivalently the twist will not be much greater then the critical
one) before the instability sets in and completely alters the system.
Therefore, even though the line-tied eigenfunction does approach the
periodic one as $L\rightarrow\infty$, the physical significance is
not clear. All these considerations point to the much more important
nonlinear problem. Our next step will be to look at the nonlinear
equilibrium after the instability, which will serve as a direct comparison
to the nonlinear equilibrium given by RDR. Presently, there is still
no consensus as to whether an infinitely thin current sheet or a current
layer with finite thickness would form in a line-tied configuration.
And we hope we can further clarify this issue.

\begin{acknowledgments}
The authors would like to acknowledge beneficial discussions with
Drs. A. Bhattacharjee, G. L. Delzanno, E. G. Evstatiev, J. M. Finn,
C. B. Forest, A. B. Hassam, C. C. Hegna, V. V. Mirnov, and C. S. Ng.
\end{acknowledgments}
\appendix

\section{Numerical Method\label{sec:Numerical-Method}}

We discretize the eigenmode equations by a Petrov-Galerkin scheme\cite{QuarteroniV1994}.
For a generalized eigenvalue problem of the form $\lambda\mathcal{L}_{1}u=\mathcal{L}_{2}u$,
the weak form of the same system is $\lambda(\mathcal{L}_{1}u,w)=(\mathcal{L}_{2}u,w)$
for all $w$ in a suitable space of test functions, where $(\cdot,\cdot)$
denotes the inner product over the spatial domain. For our numerical
method, we will approximate $u$ and $w$ with finite dimensional
spaces, therefore turning the original partial differential eigenvalue
equations into a matrix eigenvalue problem. In particular, we expand
the variables $\xi_{r}$ and $i\tilde{\xi_{\eta}}$ as \begin{equation}
\left[\begin{array}{c}
\xi_{r}\\
i\tilde{\xi_{\eta}}\end{array}\right]=\sum_{\mu=1}^{2N_{r}}\sum_{\nu=1}^{N_{z}}a_{\mu\nu}\left[\begin{array}{c}
\psi_{r}^{\mu}(r)\\
\psi_{\eta}^{\mu}(r)\end{array}\right]\phi^{\nu}(z).\label{eq:expansion}\end{equation}
 Substituting this expression into the eigenmode equations, we then
require the resulting equations to be satisfied identically when projected
onto the $2N_{r}\times N_{z}$ basis functions of the test space\begin{equation}
\left[\begin{array}{c}
\bar{\psi_{r}^{\mu}}(r)\\
\bar{\psi_{\eta}^{\mu}}(r)\end{array}\right]\bar{\phi^{\nu}}(z).\label{eq:separate}\end{equation}
For our convenience, we take \begin{equation}
\left[\begin{array}{c}
\bar{\psi_{r}^{\mu}}(r)\\
\bar{\psi_{\eta}^{\mu}}(r)\end{array}\right]=\left[\begin{array}{cc}
r & 0\\
-\frac{r^{2}V_{Az}}{mV_{A}}\left(\partial_{r}+\frac{1}{r}\right) & r\end{array}\right]\left[\begin{array}{c}
\psi_{r}^{\mu}(r)\\
\psi_{\eta}^{\mu}(r)\end{array}\right],\label{eq:dual-basis-r}\end{equation}
\begin{equation}
\bar{\phi^{\nu}}(z)=\frac{\phi^{\nu}(z)}{\sqrt{1-(2z/L)^{2}}},\label{eq:dual-basis-z}\end{equation}
and the inner product is defined as\begin{equation}
\left(\left[\begin{array}{c}
f_{1}\\
g_{1}\end{array}\right],\left[\begin{array}{c}
f_{2}\\
g_{2}\end{array}\right]\right)\equiv\int_{-L/2}^{L/2}dz\int_{0}^{r_{0}}dr(f_{1}f_{2}^{*}+g_{1}g_{2}^{*}).\label{eq:inner-product}\end{equation}
The choice of $\bar{\psi_{r}^{\mu}}$ and $\bar{\psi_{\eta}^{\mu}}$
gives the correct form of energy integration in the $r$ direction,
and the factor $1/\sqrt{1-(2z/L)^{2}}$ in $\bar{\phi^{\nu}}$ allows
us to utilize the orthogonality of Chebyshev polynomials, which are
used to construct the basis functions (see details below).

After lengthy calculations and integration by parts, the discretized
approximation to the eigenmode equation can be written as 

\begin{equation}
-\gamma^{2}M_{1}^{\alpha\beta}D_{0}^{\mu\nu}a_{\beta\nu}=\left(M_{2}^{\alpha\beta}D_{2}^{\mu\nu}+M_{3}^{\alpha\beta}D_{1}^{\mu\nu}+M_{4}^{\alpha\beta}D_{0}^{\mu\nu}\right)a_{\beta\nu},\label{eq:eig-reform}\end{equation}
where\begin{eqnarray}
M_{1}^{\alpha\beta} & = & -\int_{0}^{r_{0}}r\Bigg[A_{1}\frac{d\psi_{r}^{\alpha*}}{dr}\frac{d\psi_{r}^{\beta}}{dr}+A_{2}\psi_{r}^{\alpha*}\psi_{r}^{\beta}+A_{3}\left(\psi_{\eta}^{\alpha*}\frac{d\psi_{r}^{\beta}}{dr}+\psi_{\eta}^{\beta}\frac{d\psi_{r}^{\alpha*}}{dr}\right)\nonumber \\
 &  & +A_{4}(\psi_{\eta}^{\alpha*}\psi_{r}^{\beta}+\psi_{\eta}^{\beta}\psi_{r}^{\alpha*})+\psi_{\eta}^{\alpha*}\psi_{\eta}^{\beta}\Bigg]dr,\label{eq:M1}\end{eqnarray}
with\begin{equation}
A_{1}=\frac{r^{2}V_{Az}^{2}}{m^{2}V_{A}^{2}},\label{eq:A1}\end{equation}
\begin{equation}
A_{2}=1-\frac{1}{m^{2}V_{A}^{2}}\left(V_{Az}^{2}-2rV_{A\phi}V_{A\phi}'-2\frac{V_{A\phi}^{4}}{V_{A}^{2}}\right),\label{eq:A2}\end{equation}
\begin{equation}
A_{3}=-\frac{rV_{Az}}{mV_{A}},\label{eq:A3}\end{equation}
\begin{equation}
A_{4}=-\frac{V_{Az}}{mV_{A}};\label{eq:A4}\end{equation}
\begin{eqnarray}
M_{2}^{\alpha\beta} & = & \int_{0}^{r_{0}}\Bigg[A_{5}\frac{d\psi_{r}^{\alpha*}}{dr}\frac{d\psi_{r}^{\beta}}{dr}+A_{6}\psi_{r}^{\alpha*}\psi_{r}^{\beta}+A_{7}\left(\psi_{\eta}^{\alpha*}\frac{d\psi_{r}^{\beta}}{dr}+\psi_{\eta}^{\beta}\frac{d\psi_{r}^{\alpha*}}{dr}\right)\nonumber \\
 &  & +A_{8}(\psi_{\eta}^{\alpha*}\psi_{r}^{\beta}+\psi_{\eta}^{\beta}\psi_{r}^{\alpha*})+A_{9}\psi_{\eta}^{\alpha*}\psi_{\eta}^{\beta}\Bigg]dr\label{eq:M2}\end{eqnarray}
with\begin{equation}
A_{5}=\frac{r^{3}V_{Az}^{2}}{m^{2}},\label{eq:A5}\end{equation}
\begin{equation}
A_{6}=\frac{m^{2}-1}{m^{2}}rV_{Az}^{2}+\frac{2rV_{A\phi}^{2}}{m^{2}}+\frac{2r^{2}V_{A\phi}V_{A\phi}'}{m^{2}},\label{eq:A6}\end{equation}
\begin{equation}
A_{7}=-\frac{r^{2}V_{A}V_{Az}}{m},\label{eq:A7}\end{equation}
\begin{equation}
A_{8}=-\frac{rV_{A}V_{Az}}{m},\label{eq:A8}\end{equation}
\begin{equation}
A_{9}=rV_{A}^{2};\label{eq:A9}\end{equation}
\begin{eqnarray}
M_{3}^{\alpha\beta} & = & \int_{0}^{r_{0}}i\Bigg[A_{10}\frac{d\psi_{r}^{\alpha*}}{dr}\frac{d\psi_{r}^{\beta}}{dr}+A_{11}\psi_{r}^{\alpha*}\psi_{r}^{\beta}+A_{12}\left(\psi_{\eta}^{\alpha*}\frac{d\psi_{r}^{\beta}}{dr}+\psi_{\eta}^{\beta}\frac{d\psi_{r}^{\alpha*}}{dr}\right)\nonumber \\
 &  & +A_{13}(\psi_{\eta}^{\alpha*}\psi_{r}^{\beta}+\psi_{\eta}^{\beta}\psi_{r}^{\alpha*})\Bigg]dr,\label{eq:M3}\end{eqnarray}
with\begin{equation}
A_{10}=\frac{2r^{2}V_{Az}V_{A\phi}}{m},\label{eq:A10}\end{equation}
\begin{equation}
A_{11}=\frac{2(m^{2}-1)V_{A\phi}V_{Az}}{m},\label{eq:A11}\end{equation}
\begin{equation}
A_{12}=-rV_{A\phi}V_{A},\label{eq:A12}\end{equation}
\begin{equation}
A_{13}=V_{A}V_{A\phi};\label{eq:A13}\end{equation}
\begin{equation}
M_{4}^{\alpha\beta}=\int_{0}^{r_{0}}\left[A_{14}\frac{d\psi_{r}^{\alpha}}{dr}\frac{d\psi_{r}^{\beta}}{dr}+A_{15}\psi_{r}^{\alpha}\psi_{r}^{\beta}+A_{16}\psi_{\eta}^{\alpha}\psi_{\eta}^{\beta}\right]dr,\label{eq:M4}\end{equation}
with\begin{equation}
A_{14}=-rV_{A\phi}^{2},\label{eq:A14}\end{equation}
\begin{equation}
A_{15}=\frac{(1-m^{2})V_{A\phi}^{2}}{r},\label{eq:A15}\end{equation}
\begin{equation}
A_{16}=-\frac{m^{2}V_{A}^{2}}{r};\label{eq:A16}\end{equation}
and \begin{equation}
D_{2}^{\mu\nu}=\int_{-L/2}^{L/2}\bar{\phi^{\mu}}^{*}\frac{d^{2}\phi^{\nu}}{dz}dz,\label{eq:D2}\end{equation}
\begin{equation}
D_{1}^{\mu\nu}=\int_{-L/2}^{L/2}\bar{\phi^{\mu}}^{*}\frac{d\phi^{\nu}}{dz}dz,\label{eq:D1}\end{equation}
\begin{equation}
D_{0}^{\mu\nu}=\int_{-L/2}^{L/2}\bar{\phi^{\mu}}^{*}\phi^{\nu}dz.\label{eq:D0}\end{equation}

We have tried various basis functions. A straightforward choice satisfying
boundary conditions and regularity conditions at $r=0$ based on Chebyshev
polynomials would be \begin{eqnarray}
\left[\begin{array}{c}
\psi_{r}^{\mu}(r)\\
\psi_{\eta}^{\mu}(r)\end{array}\right] & = & \left[\begin{array}{c}
(r/r_{0})^{|m|-1}[T_{\mu}(2r/r_{0}-1)-1]\\
0\end{array}\right]\,\,\,\,\,\text{when }\mu\le N_{r},\label{eq:basis-r}\end{eqnarray}
\begin{equation}
\left[\begin{array}{c}
\psi_{r}^{\mu}(r)\\
\psi_{\eta}^{\mu}(r)\end{array}\right]=\left[\begin{array}{c}
0\\
(r/r_{0})^{|m|+1}T_{\mu-N_{r}-1}(2r/r_{0}-1)\end{array}\right]\,\,\,\,\text{when }\mu>N_{r},\label{eq:basis-r1}\end{equation}
and\begin{equation}
\phi^{\nu}(z)=T_{\nu+1}(2z/L)-T_{\nu-1}(2z/L),\label{eq:basis-z}\end{equation}
where $T_{n}$ denotes the Chebyshev polynomial of order $n$.\cite{Boyd2001}
This choice of basis functions works well, but a significant amount
of basis functions are needed, as the eigenmode typically possesses
a steep gradient at some radius $r=s$ and many oscillations along
$z$. To remedy the steep gradient problem, we need to pack more resolution
about $r=s$. This is implemented by successively applying three arctan/tan
mappings, which are recommended in Ref. \cite{Boyd2001}, as follows:
First we map $r\in[0,r_{0}]\rightarrow r_{1}\in[0,1]$ with $s\rightarrow1/2$
by\begin{equation}
r_{1}=\frac{2}{\pi}\tan^{-1}\left(a_{1}\tan\left(\frac{\pi}{2}\frac{r}{r_{o}}\right)\right),\label{eq:r1}\end{equation}
with \begin{equation}
a_{1}=\frac{1}{\tan\left(\frac{\pi}{2}\frac{s}{r_{0}}\right)}.\label{eq:a1}\end{equation}
Then a second mapping from $[0,1]\rightarrow[0,1]$\begin{equation}
r_{2}=\frac{1}{\pi}\tan^{-1}\left(a_{2}\tan\left(\pi\left(r_{1}-\frac{1}{2}\right)\right)\right)+\frac{1}{2}\label{eq:r2}\end{equation}
is applied, which packs more resolution about $r_{1}=1/2$ (which
is mapped into $r_{2}=1/2$), with larger $a_{2}$ for more packing.
Finally, we map $r_{2}\in[0,1]\rightarrow r_{3}\in[0,1]$ with $1/2\rightarrow s'$
by \begin{equation}
r_{3}=\frac{2}{\pi}\tan^{-1}\left(a_{3}\tan\left(\frac{\pi}{2}r\right)\right),\label{eq:r3}\end{equation}
with\begin{equation}
a_{3}=\tan\left(\frac{\pi}{2}s'\right).\label{eq:a3}\end{equation}
In this mapping, $s'$ can be any number between $0$ and $1$, but
we usually choose some $s'>1/2$. This is motivated by the observation
that typically the eigenmode structure is rather trivial ($\simeq0$)
in the region $r>s$, therefore one can put less resolution there.
The basis functions are then defined in terms of $r_{3}$ as \begin{eqnarray}
\left[\begin{array}{c}
\psi_{r}^{\mu}\\
\psi_{\eta}^{\mu}\end{array}\right] & = & \left[\begin{array}{c}
(r_{3})^{|m|-1}[T_{\mu}(2r_{3}-1)-1]\\
0\end{array}\right]\,\,\,\,\,\text{when }\mu\le N_{r},\label{eq:basis-r-2}\end{eqnarray}
and\begin{equation}
\left[\begin{array}{c}
\psi_{r}^{\mu}\\
\psi_{\eta}^{\mu}\end{array}\right]=\left[\begin{array}{c}
0\\
(r_{3})^{|m|+1}T_{\mu-N_{r}-1}(2r_{3}-1)\end{array}\right]\,\,\,\,\text{when }\mu>N_{r}.\label{eq:basis-r3}\end{equation}
To resolve the many oscillations along z without too many basis functions,
we use the observation that the fastest growing mode, upon dividing
by $e^{ik_{0}z}$ with $k_{0}$ being the wave number of the fastest
growing mode in the corresponding infinite system, yields a smooth
function with length scales $\sim L$. This suggests that one may
use \begin{equation}
\phi^{\nu}(z)=e^{ik_{0}z}(T_{\nu+1}(2z/L)-T_{\nu-1}(2z/L))\label{eq:basis-z1}\end{equation}
 as the basis functions. This proves to be very efficient.

The integration along $r$ is carried out numerically by Gaussian
quadrature. The integration along $z$ can be done analytically,\cite{MasonH2002}
which gives\begin{equation}
D_{0}^{\mu\nu}=\frac{L\pi}{2}\left[\begin{array}{cccccc}
3/2 & 0 & -1/2\\
0 & 1 & 0 & -1/2\\
-1/2 & 0 & 1 & 0 & \ddots\\
 & -1/2 & 0 & \ddots & \ddots\\
 &  & \ddots & \ddots\end{array}\right],\label{eq:D0a}\end{equation}
\begin{equation}
D_{1}^{\mu\nu}=\pi\left[\begin{array}{cccccc}
0 & 1\\
-2 & 0 & 2\\
 & -3 & 0 & 3\\
 &  & -4 & 0 & \ddots\\
 &  &  & \ddots & \ddots\end{array}\right]+ik_{0}D_{0}^{\mu\nu},\label{eq:D1a}\end{equation}
\begin{equation}
D_{2}^{\mu\nu}=\frac{2\pi}{L}\times\left\{ \begin{array}{cc}
-2\mu(\mu+1)\,\,\,\, & \mu=\nu\\
-4\mu\,\,\,\, & \text{$\mu<\nu$, $\nu-\mu$ even}\end{array}\right.+2ik_{0}\pi\left[\begin{array}{cccccc}
0 & 1\\
-2 & 0 & 2\\
 & -3 & 0 & 3\\
 &  & -4 & 0 & \ddots\\
 &  &  & \ddots & \ddots\end{array}\right]-k_{0}^{2}D_{0}^{\mu\nu}.\label{eq:D2a}\end{equation}
The resulting finite dimensional eigenvalue problem is then solved
by a Jacobi-Davidson iteration method, which is described in detail
in Ref. \cite{SleijpenV1996,FokkemaSV1998}. 

The growth rates of unstable modes are benchmarked with an initial-value
MHD code, NIMROD;\cite{SovinecGGBNKSPTCN2004} and the critical lengths
for instability are also in good agreement with the results published
in Ref. \cite{VanderLindenH1999}. A periodic version of the code
is also developed, in which spatio-temporal dependence of the form
$\exp(ikz+im\phi+\gamma t)$ is assumed.

\section{Semi-Analytic Analysis of Line-tied Internal Kink\label{sec:semi-analytic}}

The linear analysis of the periodic internal kink modes in RDR goes
as follows. First the eigenvalue problem is approximated by a single
equation for $\xi_{r}$. The equation is then solved approximately
in the internal layer and the outer region separately. Finally, asymptotic
matching of the inner and the outer solutions gives the growth rate
of the unstable mode. In this appendix, we try to extend the analysis
to a line-tied system. An estimate for the critical length and the
relation between the growth rate and the system length are derived.
It should be pointed out though, that the analysis is not completely
satisfactory. At certain point, we are forced to appeal to numerical
work to validate certain steps. That is why we call it semi-analytic.

Following the steps of the analysis of RDR, first we have to derive
a single equation for $\xi_{r}$. We start with the variational principle
for the eigenvalue problem.\cite{Freidberg1987} Assuming $m=\pm1$,
for a displacement $\bm{\xi}$, the energy variation can be written
in term of $\xi_{r}$ and $\tilde{\xi_{\eta}}$ as \begin{eqnarray}
W(\bm{\xi}) & = & \int_{-L/2}^{L/2}dz\int_{0}^{r_{0}}dr\Bigg[r^{3}\left|\mathbf{V}_{A}\cdot\nabla\xi_{r}'\right|^{2}+A_{6}\left|\partial_{z}\xi_{r}\right|^{2}-iA_{7}\left(\partial_{z}\tilde{\xi_{\eta}}^{*}\partial_{z}\xi_{r}'-c.c.\right)\nonumber \\
 &  & -iA_{8}\left(\partial_{z}\tilde{\xi_{\eta}}^{*}\partial_{z}\xi_{r}-c.c.\right)+A_{9}\left|\partial_{z}\tilde{\xi_{\eta}}\right|^{2}\nonumber \\
 &  & -A_{12}\left(\tilde{\xi_{\eta}}^{*}\partial_{z}\xi_{r}'+c.c.\right)-A_{13}\left(\tilde{\xi_{\eta}}^{*}\partial_{z}\xi_{r}+c.c.\right)-A_{16}\left|\tilde{\xi_{\eta}}\right|^{2}\Bigg]\label{eq:energy1}\end{eqnarray}
and the inertia term can be written as \begin{eqnarray}
I(\bm{\xi}) & = & \int_{-L/2}^{L/2}dz\int_{0}^{r_{0}}dr\, r\Bigg[A_{1}\left|\xi_{r}'\right|^{2}+A_{2}\left|\xi_{r}\right|^{2}-iA_{3}\left(\tilde{\xi_{\eta}}^{*}\xi_{r}'-c.c.\right)\nonumber \\
 &  & -iA_{4}\left(\tilde{\xi_{\eta}}^{*}\xi_{r}-c.c.\right)+\left|\tilde{\xi_{\eta}}\right|^{2}\Bigg],\label{eq:inertia}\end{eqnarray}
where the coefficients $A_{1}$---$A_{16}$ are defined in Appendix
\ref{sec:Numerical-Method}, and $c.c.$ denotes complex conjugate.
The eigenvalue problem is equivalent to the following variational
problem:\cite{Freidberg1987}\begin{equation}
\gamma^{2}=-\frac{W(\bm{\xi})}{I(\bm{\xi})},\label{eq:energy}\end{equation}
and \begin{equation}
\delta\frac{W(\bm{\xi})}{I(\bm{\xi})}=0.\label{eq:variation}\end{equation}
Since $\tilde{\xi_{\eta}}$ is small for kink modes in the small $\epsilon$
limit, we may approximate $I(\bm{\xi})$ as \begin{equation}
I(\bm{\xi})\simeq\int_{-L/2}^{L/2}dz\int_{0}^{r_{0}}dr\, r^{3}\left|\xi_{r}'\right|^{2}.\label{eq:approximate-inertia}\end{equation}
That is, that $I(\bm{\xi})$ only weakly depends on $\tilde{\xi_{\eta}}$
; therefore we may minimize $W(\bm{\xi})$ with respect to $\tilde{\xi_{\eta}}$
first in the variational problem (\ref{eq:variation}). The minimization
gives the following relation between $\xi_{r}$ and $\tilde{\xi_{\eta}}$:\begin{equation}
(A_{9}\partial_{z}^{2}+A_{16})\tilde{\xi_{\eta}}=iA_{7}\partial_{z}^{2}\xi_{r}'+iA_{8}\partial_{z}^{2}\xi_{r}-A_{12}\partial_{z}\xi_{r}'-A_{13}\partial_{z}\xi_{r}.\label{eq:eta-minimize}\end{equation}
Since $r\partial_{z}\sim ikr\sim\epsilon$, this shows that $\tilde{\xi_{\eta}}\sim\epsilon^{2}$;
therefore the approximation (\ref{eq:approximate-inertia}) is self-consistent.
Using Eq. (\ref{eq:eta-minimize}) in (\ref{eq:energy}) to eliminate
as many $\tilde{\xi_{\eta}}$ as possible, we may write $W$ as \begin{eqnarray}
W(\bm{\xi}) & = & \int_{-L/2}^{L/2}dz\int_{0}^{r_{0}}dr\left(r^{3}\left|\mathbf{V}_{A}\cdot\nabla\xi_{r}'\right|^{2}+A_{6}\left|\partial_{z}\xi_{r}\right|^{2}-A_{9}\left|\partial_{z}\tilde{\xi_{\eta}}\right|^{2}+A_{16}\left|\tilde{\xi_{\eta}}\right|^{2}\right).\label{eq:WW}\end{eqnarray}
To complete the minimization, we still have to solve (\ref{eq:eta-minimize})
for $\tilde{\xi_{\eta}}$ in terms of $\xi_{r}$. This could formally
be done in terms of the Green's function, but let us proceed with
the following approximation. We expand $W$ up to $O(\epsilon^{4})$.
The term $A_{9}\left|\partial_{z}\tilde{\xi_{\eta}}\right|^{2}$ in
the integrand is $O(\epsilon^{6})$ therefore is negligible; and from
Eq. (\ref{eq:eta-minimize}), \begin{equation}
\tilde{\xi_{\eta}}=\left(iA_{7}\partial_{z}^{2}\xi_{r}'+iA_{8}\partial_{z}^{2}\xi_{r}-A_{12}\partial_{z}\xi_{r}'-A_{13}\partial_{z}\xi_{r}\right)/A_{16}+O(\epsilon^{4}).\label{eq:eta-approx}\end{equation}
After some algebra and integration by parts in $r$, we have \begin{eqnarray}
W & = & \int_{-L/2}^{L/2}dz\int_{0}^{r_{0}}dr\bigg[r^{3}\left|\mathbf{V}_{A}\cdot\nabla\xi_{r}'\right|^{2}-r^{5}\left|\partial_{z}(\mathbf{V}_{A}\cdot\nabla\xi_{r}')\right|^{2}+3r^{3}V_{Az}^{2}\left|\partial_{z}^{2}\xi_{r}\right|^{2}-rV_{A\phi}^{2}\left|\partial_{z}\xi_{r}\right|^{2}\nonumber \\
 &  & +imV_{Az}V_{A\phi}r^{2}\left(r\partial_{z}\xi_{r}'^{*}\partial_{z}^{2}\xi_{r}-r\partial_{z}\xi_{r}^{*}\partial_{z}^{2}\xi_{r}'+\partial_{z}^{2}\xi_{r}^{*}\partial_{z}\xi_{r}-c.c.\right)\bigg]+O(\epsilon^{6}).\label{eq:approx-energy1}\end{eqnarray}
Integration by parts in $z$ gives\begin{multline}
\int_{-L/2}^{L/2}dz\int_{0}^{r_{0}}dr\, imV_{Az}V_{A\phi}r^{2}\left(r\partial_{z}\xi_{r}'^{*}\partial_{z}^{2}\xi_{r}-r\partial_{z}\xi_{r}^{*}\partial_{z}^{2}\xi_{r}'-c.c.\right)\\
=\int_{0}^{r_{0}}dr\, imV_{Az}V_{A\phi}r^{2}\left(r\partial_{z}\xi_{r}'^{*}\partial_{z}\xi_{r}-c.c.\right)_{-L/2}^{L/2}.\label{eq:byparts}\end{multline}
This term may be negligible in Eq. (\ref{eq:approx-energy1}), as
one can estimate \begin{equation}
\left(r\partial_{z}\xi_{r}'^{*}\partial_{z}\xi_{r}-c.c.\right)_{-L/2}^{L/2}:\int_{-L/2}^{L/2}dz\left(\partial_{z}^{2}\xi_{r}^{*}\partial_{z}\xi_{r}-c.c.\right)\sim1:kL\sim\epsilon^{2}:1,\label{eq:term-estimate}\end{equation}
where in the last step we use the observed scaling law that to have
unstable mode, $L>L_{c}\propto1/\epsilon^{3}$. We may also neglect
the second term in the integrand of Eq. (\ref{eq:approx-energy1}),
since \begin{equation}
r^{5}\left|\partial_{z}(\mathbf{V}_{A}\cdot\nabla\xi_{r}')\right|^{2}:r^{3}\left|\mathbf{V}_{A}\cdot\nabla\xi_{r}'\right|^{2}\sim k^{2}r^{2}:1\sim\epsilon^{2}:1.\label{eq:term-estimate1}\end{equation}
Therefore $W$ can be further simplified to \begin{eqnarray}
W & \simeq & \int_{-L/2}^{L/2}dz\int_{0}^{r_{0}}dr\bigg[r^{3}\left|\mathbf{V}_{A}\cdot\nabla\xi_{r}'\right|^{2}+3r^{3}V_{Az}^{2}\left|\partial_{z}^{2}\xi_{r}\right|^{2}-rV_{A\phi}^{2}\left|\partial_{z}\xi_{r}\right|^{2}\nonumber \\
 &  & +imV_{Az}V_{A\phi}r^{2}\left(\partial_{z}^{2}\xi_{r}^{*}\partial_{z}\xi_{r}-c.c.\right)\bigg].\label{eq:approx-energy2}\end{eqnarray}
Now we have the expressions for the energy $W$ and the inertia $I$
in terms of $\xi_{r}$ alone, Eqs. (\ref{eq:approx-energy2}) and
(\ref{eq:approximate-inertia}). We may now substitute the expressions
into the variational principle (\ref{eq:variation}) to obtain an
single partial differential equation which approximates the full eigenvalue
problem. There is a subtle point here. The form of the approximate
$W$ requires four boundary conditions for $\xi_{r}$ in the $z$
direction. Two apparent ones are the line-tied boundary conditions
$\xi_{r}|_{z=\pm L/2}=0$. A possible choice of another two boundary
conditions is $\partial_{z}\xi_{r}|_{z=\pm L/2}=0$, which makes all
boundary terms vanish when turning Eq. (\ref{eq:variation}) into
a partial differential equation through integration by parts. \emph{Assuming}
that, the approximate eigenvalue equation is \begin{equation}
\left(r^{3}(-\gamma^{2}+(\mathbf{V}_{A}\cdot\nabla)^{2})\xi_{r}'\right)'+\left(3r^{3}V_{Az}^{2}\partial_{z}^{4}+2imr^{2}V_{Az}V_{A\phi}\partial_{z}^{3}+rV_{A\phi}^{2}\partial_{z}^{2}\right)\xi_{r}=0.\label{eq:eig-approx1}\end{equation}
It is hard to justify this particular choice of boundary conditions
rigorously. In particular, it is not the same as requiring $\tilde{\xi_{\eta}}=0$
at the boundary (see Eq. (\ref{eq:eta-approx})). Nonetheless, numerical
solutions show that this new eigenvalue problem is indeed a good approximation
to the original one in the small $\epsilon$ limit. As we will see,
it turns out that $\xi_{r}$ is vanishingly small within the boundary
layers near $z=\pm L/2$, therefore indeed $\partial_{z}\xi_{r}|_{z=\pm L/2}\simeq0$. 

One nice feature of Eq. (\ref{eq:eig-approx1}) is that, if we consider
a periodic problem by letting $\partial_{z}\rightarrow ik$, the approximate
equation of RDR is recovered. This allows us to apply a similar asymptotic
technique to the present problem. Motivated by the numerical solution,
let us write the eigenfunction as $\xi_{r}(r,z)=h(r,z)e^{ik_{0}z}$,
where $k_{0}$ is the wavenumber of the fastest growing periodic mode,
$e^{ik_{0}z}$ is the fast oscillating part, and $h(r,z)$ is the
envelope slowly varying along the $z$ direction. Substituting this
into the energy Eq. (\ref{eq:approx-energy2}), the dominant term
in the integrand is the positive definite $r^{3}\left|\mathbf{V}_{A}\cdot\nabla\xi_{r}'\right|^{2}=r^{3}\left|\left(V_{Az}\partial_{z}+i\mathbf{k}_{0}\cdot\mathbf{V}_{A}\right)h'\right|^{2}\sim O(\epsilon^{2}),$
and all other terms are $O(\epsilon^{4})$. For an unstable mode,
which requires $W<0$, we must have $h'\simeq0$ everywhere except
where $\mathbf{k}_{0}\cdot\mathbf{V}_{A}\simeq0$. Therefore, the
internal layer occurs at $r=r_{s}$, where $\mathbf{k}_{0}\cdot\mathbf{V}_{A}=0$;
outside of the internal layer, $h(r,z)\simeq h_{0}(z)$ when $r<r_{s}$,
and $h(r,z)\simeq0$ when $r>r_{s}$. 

\begin{figure}
\includegraphics[scale=0.5]{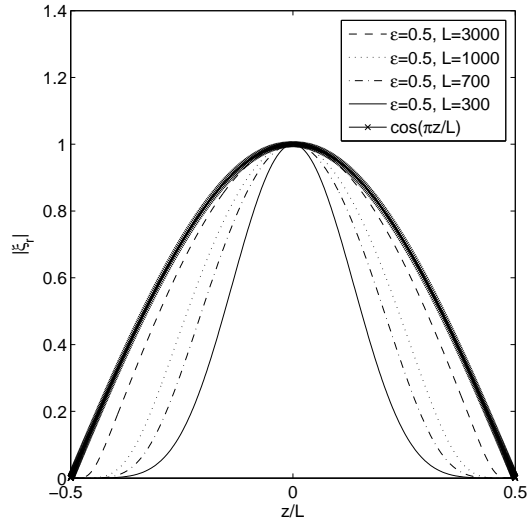}

\caption{Envelope of $\xi_{r}$ at $r=0$ for the case $\epsilon=0.5$ with
various $L$. The envelope function is confined to the central region
and is vanishingly small within the boundary layers near the ends.
The envelope function approaches $\cos(\pi z/L)$, which is shown
as a reference, in the limit $L\rightarrow\infty$. \label{cap:zcut}}
\end{figure}

Integrating Eq. (\ref{eq:eig-approx1}) once along $r$, we have\begin{multline}
\left(-\gamma^{2}+(\mathbf{V}_{A}\cdot\nabla)^{2}\right)\xi_{r}'(r,z)=\\
-\frac{1}{r^{3}}\int_{0}^{r}d\bar{r}\left(3\bar{r}^{3}V_{Az}^{2}(\bar{r})\partial_{z}^{4}+2im\bar{r}^{2}V_{Az}(\bar{r})V_{A\phi}(\bar{r})\partial_{z}^{3}+\bar{r}V_{A\phi}^{2}(\bar{r})\partial_{z}^{2}\right)\xi_{r}(\bar{r},z).\label{eq:eig-approx2}\end{multline}
 Using $\xi_{r}(r,z)=h(r,z)e^{ik_{0}z}$ in Eq. (\ref{eq:eig-approx2}),
in the region $r<r_{s}$ we have\begin{equation}
\left(-\gamma^{2}+\left(V_{Az}\partial_{z}+i\mathbf{k}_{0}\cdot\mathbf{V}_{A}\right)^{2}\right)h'\simeq\frac{-h_{0}(z)}{r^{3}}\int_{0}^{r}g_{1}(\bar{r})d\bar{r},\label{eq:outer1}\end{equation}
where \begin{equation}
g_{1}(r)=3k_{0}^{4}r^{3}V_{Az}^{2}+2mk_{0}^{3}r^{2}V_{Az}V_{A\phi}-k_{0}^{2}rV_{A\phi}^{2}.\label{eq:g1-def}\end{equation}
In the RHS of Eq. (\ref{eq:outer1}) we make use of that $h(r,z)\simeq h_{0}(z)$
when $r<r_{s}$, and all derivatives on the envelope function are
neglected as they are smaller compared to derivatives on the fast
oscillating part $e^{ik_{0}z}$. Within the internal layer, we may
assume that $\left(-\gamma^{2}+\left(V_{Az}\partial_{z}+i\mathbf{k}_{0}\cdot\mathbf{V}_{A}\right)^{2}\right)h'$
is nearly independent of $r$; that is \begin{equation}
\left(-\gamma^{2}+\left(V_{Az}\partial_{z}+i\mathbf{k}_{0}\cdot\mathbf{V}_{A}\right)^{2}\right)h'=f(z).\label{eq:inner}\end{equation}
The function $f(z)$ can be determined from the outer region by evaluating
Eq. (\ref{eq:outer1}) at $r=r_{s}$:\begin{equation}
f(z)=\frac{-h_{0}(z)}{r_{s}^{3}}\int_{0}^{r_{s}}g_{1}(r)dr.\label{eq:f}\end{equation}
Furthermore, $V_{Az}$ can be approximated by $V_{Az}(r_{s})$, and
$\mathbf{k}_{0}\cdot\mathbf{V}_{A}\simeq(\mathbf{k}_{0}\cdot\mathbf{V}_{A})'|_{r_{s}}x\simeq-V_{Az}(r_{s})\zeta'(r_{s})x$,
where $x\equiv r-r_{s}$ and \begin{equation}
\zeta(r)\equiv-\frac{mV_{A\phi}}{rV_{Az}}.\label{eq:q}\end{equation}
Using these approximations, the governing equation within the internal
layer is \begin{equation}
\left(V_{Az}^{2}(\partial_{z}-i\zeta'x)^{2}-\gamma^{2}\right)h'=f(z).\label{eq:inner1}\end{equation}
Here $V_{Az}$ and $\zeta'$ are evaluated at $r=r_{s}$. Eq. (\ref{eq:inner1})
can be formally solved using the Green's function:\begin{equation}
G(x;z,\bar{z})=\frac{e^{i\zeta'x(z-\bar{z})}}{\gamma V_{Az}}\frac{\sinh\left(\gamma(z_{>}-L/2)/V_{Az}\right)\sinh\left(\gamma(z_{<}+L/2)/V_{Az}\right)}{\sinh\left(\gamma L/V_{Az}\right)},\label{eq:green}\end{equation}
where $z_{>}$ ($z_{<}$) is the larger (smaller) one of $z$ and
$\bar{z}$. The solution is \begin{equation}
h'(x,z)=\int_{-L/2}^{L/2}G(x;z,\bar{z})f(\bar{z})d\bar{z}\label{eq:h-sol}\end{equation}
The solution in the inner region has to match the solution in the
outer region asymptotically. That requires that the jump in the inner
solution as $x$ goes from $-\infty$ to $\infty$ be equal to the
jump in the outer solution from the region $r<r_{s}$ to the region
$r>r_{s}$: \begin{equation}
\int_{-\infty}^{\infty}dx\, h'(x,z)=-h_{0}(z).\label{eq:matching}\end{equation}
The only $x$ dependence in $h'(x,z)$ comes from the factor $e^{i\zeta'x(z-\bar{z})}$
in the Green's function, and the integration can be done easily:\begin{equation}
\int_{-\infty}^{\infty}e^{i\zeta'x(z-\bar{z})}dx=\frac{2\pi}{\left|\zeta'\right|}\delta(z-\bar{z}).\label{eq:integral}\end{equation}
Using Eqs. (\ref{eq:integral}), (\ref{eq:h-sol}), (\ref{eq:green}),
and (\ref{eq:f}) in (\ref{eq:matching}), the asymptotic matching
condition is \begin{equation}
\frac{2\pi}{\left|\zeta'\right|}\frac{\sinh\left(\gamma(z-L/2)/V_{Az}\right)\sinh\left(\gamma(z+L/2)/V_{Az}\right)}{\gamma V_{Az}\sinh\left(\gamma L/V_{Az}\right)}\frac{\int_{0}^{r_{s}}g_{1}(r)dr}{r_{s}^{3}}h_{0}(z)=h_{0}(z).\label{eq:matching1}\end{equation}
It seems we are now in trouble. If we eliminate $h_{0}(z)$ from both
sides, we end up requiring that a function of $z$ equals to unity,
which is impossible. However, notice that in the limit $\gamma L/V_{Az}\gg1$,
we have \begin{equation}
\frac{\sinh\left(\gamma(z-L/2)/V_{Az}\right)\sinh\left(\gamma(z+L/2)/V_{Az}\right)}{\gamma V_{Az}\sinh\left(\gamma L/V_{Az}\right)}\simeq-\frac{1}{2\gamma\left|V_{Az}\right|}\label{eq:long-limit}\end{equation}
for most of the domain $z\in[-L/2,L/2]$, except in the two boundary
layers of thickness $\sim V_{Az}/\gamma$ near both ends, where it
drops to zero quickly. Therefore, as long as $h_{0}(z)$ is confined
to the central region outside of the boundary layers, asymptotic matching
is possible. Indeed, this qualitatively agrees with the observation
on numerical solutions, exemplified in Fig. \ref{cap:zcut} for the
case $\epsilon=0.5$ with different $L$. On the other hand, at marginal
stability, $\gamma\rightarrow0$, and \begin{equation}
\frac{\sinh\left(\gamma(z-L/2)/V_{Az}\right)\sinh\left(\gamma(z+L/2)/V_{Az}\right)}{\gamma V_{Az}\sinh\left(\gamma L/V_{Az}\right)}\rightarrow\frac{\left(z-L/2\right)\left(z+L/2\right)}{LV_{Az}^{2}}.\label{eq:marginal}\end{equation}
In this limit, the {}``boundary layers'' cover the whole domain,
and we conclude that the marginal mode must be highly localized to
the center. This is also in agreement with the numerical solutions. 

Based on these observations, it may not be unreasonable to eliminate
$h_{0}(z)$ from both side of Eq. (\ref{eq:matching1}), then approximate
the left-hand side by its value at $z=0$. After some algebra, that
gives the relation between $\gamma$ and $L$: \begin{equation}
-\frac{\pi}{\left|\zeta'V_{Az}\right|_{r_{s}}}\frac{\tanh\left(\gamma L/2\left|V_{Az}\right|_{r_{s}}\right)}{\gamma}\frac{\int_{0}^{r_{s}}g_{1}(r)dr}{r_{s}^{3}}=1.\label{eq:gamma-L}\end{equation}
Let us consider some limiting cases of Eq. (\ref{eq:gamma-L}) . In
the limit $L\rightarrow\infty$,\begin{equation}
\gamma=-\frac{\pi}{\left|\zeta'V_{Az}\right|_{r_{s}}r_{s}^{3}}\int_{0}^{r_{s}}g_{1}(r)dr\equiv\gamma_{0}.\label{eq:gamma0}\end{equation}
This essentially recovers the approximate periodic growth rate (\ref{eq:gamma-RDR})
since $(\mathbf{k}_{0}\cdot\mathbf{V}_{A})'|_{r_{s}}\simeq-V_{Az}(r_{s})\zeta'(r_{s})$.
For marginal stability, let $\gamma\rightarrow0$, we get an expression
for the critical length\begin{equation}
L_{c}=\frac{2\left|V_{Az}\right|_{r_{s}}}{\gamma_{0}}.\label{eq:Lc}\end{equation}
Finally, we can rewrite Eq. (\ref{eq:gamma-L}) in terms of $\gamma_{0}$
and $L_{c}$:\begin{equation}
\frac{\gamma_{0}}{\gamma}\tanh\left(\frac{\gamma}{\gamma_{0}}\frac{L}{L_{c}}\right)=1.\label{eq:gamma-L1}\end{equation}

In this appendix, we get good approximations for the critical length
$L_{c}$ and the relation between $\gamma$ and $L$ \emph{without}
knowing the eigenmode structure along $z$. We also have no information
about the thickness of the internal layer. This is not too surprising,
as it is well known that getting a good approximation for the eigenfunction
is more difficult than getting a good approximation for the eigenvalue.
This can be understood from the variational principle, Eq. (\ref{eq:variation}):
Any deviation of the trial function from the true eigenfunction only
leads to an error of second order in the eigenvalue. We learn from
numerical solutions that the mode structure along $z$ and the internal
layer thickness are correlated --- the more a mode is localized in
$z$, the thicker the internal layer is --- therefore they must be
solved all together. Attempts on extending the present analysis to
resolve the mode structure have been unsuccessful. We hope this appendix
can serve as a starting point for a more thorough analysis in the
future.

\bibliographystyle{apsrev}
\bibliography{ref}

\begin{table}[t]
\begin{tabular}{|c|c|c|c|c|c|c|c|c|}
\hline 
$\epsilon$&
$k_{0}$&
$\gamma_{0}$&
$\Delta_{0}$&
$\left|k_{0}\right|/\epsilon$&
$\gamma_{0}/\epsilon^{3}$ &
$\Delta_{0}/\epsilon^{2}$ &
$\gamma_{0}^{RDR}$&
$\Delta_{0}^{RDR}$\tabularnewline
\hline
\hline 
1&
-0.5287&
$8.311\times10^{-2}$&
0.3908&
0.5287&
0.083&
0.3909&
$1.203\times10^{-1}$&
0.4945\tabularnewline
\hline 
0.5&
-0.3121&
$1.546\times10^{-2}$&
0.1217&
0.6242&
0.124&
0.4868&
$1.772\times10^{-2}$&
0.1305\tabularnewline
\hline 
0.3&
-0.1970&
$3.731\times10^{-3}$&
0.04458&
0.6567&
0.138&
0.4953&
$3.939\times10^{-3}$&
0.04569\tabularnewline
\hline 
0.25&
-0.1658&
$2.204\times10^{-3}$&
0.03091&
0.6632&
0.141&
0.4946&
$2.290\times10^{-3}$&
0.03152\tabularnewline
\hline 
0.2&
-0.1337&
$1.148\times10^{-3}$&
0.01977&
0.6685&
0.144&
0.4943&
$1.177\times10^{-3}$&
0.02005\tabularnewline
\hline 
0.15&
-0.1009&
$4.908\times10^{-4}$&
0.01111&
0.6727&
0.145&
0.4938&
$4.978\times10^{-4}$&
0.01122\tabularnewline
\hline
\end{tabular}

\caption{Summary of the wave number $k_{0}$, the growth rate $\gamma_{0}$,
and the internal layer thickness $\Delta_{0}$ of the fastest growing
mode in a periodic system. For smaller $\epsilon$, these quantities
scale as expected. As a reference, $\gamma_{0}^{RDR}$ and $\Delta_{0}^{RDR}$
are the growth rate and thickness calculated by the Rosenbluth-Dagazian-Rutherford
theory, Eqs. (\ref{eq:gamma-RDR}) and (\ref{eq:internal-layer-RDR}).\label{cap:Summary-table}}
\end{table}

\begin{table}[t]
\begin{tabular}{|c|c|c|c|c|c|c|c|c|}
\hline 
$\epsilon$&
$L_{c}$&
$\Delta_{c}$&
$l_{c}$&
$l_{c}/L_{c}$&
$(\Delta_{c}-\Delta_{0})/\Delta_{0}$&
$L_{c}\epsilon^{3}$&
$\epsilon(\Delta_{c}-\Delta_{0})/\Delta_{0}$&
$l_{c}/L_{c}\epsilon$\tabularnewline
\hline
\hline 
1&
31.17&
0.5605&
15.64&
0.502&
0.4342&
31.17&
0.434&
0.502\tabularnewline
\hline 
0.5&
142.4&
0.3304&
44.97&
0.316&
1.715&
17.8&
0.858&
0.632\tabularnewline
\hline 
0.3&
556.8&
0.1842&
115.9&
0.208&
3.132&
15.03&
0.940&
0.694\tabularnewline
\hline 
0.25&
932.2&
0.1502&
164.8&
0.177&
3.859&
14.57&
0.965&
0.707\tabularnewline
\hline 
0.2&
1773&
0.1153&
262.4&
0.148&
4.832&
14.18&
0.966&
0.740\tabularnewline
\hline 
0.15&
4116&
0.08371&
468.5&
0.114&
6.535&
13.89&
0.980&
0.759\tabularnewline
\hline
\end{tabular}

\caption{Summary of $L_{c}$, $\Delta_{c}$, and $l_{c}$ for various $\epsilon$,
as well as the scaling laws they follow.\label{cap:marginal-scaling}}
\end{table}

\clearpage

\end{document}